\magnification=1200
\hoffset=.0cm
\voffset=.0cm
\baselineskip=.55cm plus .55mm minus .55mm

%
%
\def\ref#1{\lbrack#1\rbrack}
%
%
%
%
\input amssym.def
\input amssym.tex
%
%
\font\teneusm=eusm10                                   
\font\seveneusm=eusm7                   
\font\fiveeusm=eusm5                 
%
%

%
%
\font\cps=cmcsc10
%
%
\newfam\eusmfam
\textfont\eusmfam=\teneusm
\scriptfont\eusmfam=\seveneusm
\scriptscriptfont\eusmfam=\fiveeusm

\def\sh#1{\hbox{\teneusm #1}}

\def\proclaim #1. #2\par{\medbreak{\cps #1.\enspace}{\it #2}\par\medbreak}
%
%
%
%

\def\id{\hskip 1pt{\rm id}\hskip 1pt}

\def\ad{\hskip 1pt{\rm ad}\hskip 1pt}

\def\Lie{\hskip 1pt{\rm Lie}\hskip 1pt}

\def\Pois{\hskip 1pt{\rm Pois}\hskip 1pt}
\def\Ham{\hskip 1pt{\rm Ham}\hskip 1pt}
\def\Cas{\hskip 1pt{\rm Cas}\hskip 1pt}
\def\Fun{\hskip 1pt{\rm Fun}\hskip 1pt}
\def\Vect{\hskip 1pt{\rm Vect}\hskip 1pt}
\def\hst1{\hskip 1pt}

%
%
%
%

\hrule\vskip.5cm
\hbox to 16.5 truecm{April 2002 \hfil DFUB 02--03}
\hbox to 16.5 truecm{Version 1  \hfil math-ph/0205006}
\vskip.5cm\hrule
\vskip.9cm
\centerline{\bf TARGET SPACE EQUIVARIANT COHOMOLOGICAL}   
\centerline{\bf STRUCTURE OF THE POISSON SIGMA MODEL}   
\vskip.4cm
\centerline{by}
\vskip.4cm
\centerline{\bf Roberto Zucchini}
\centerline{\it Dipartimento di Fisica, Universit\`a degli Studi di Bologna}
\centerline{\it V. Irnerio 46, I-40126 Bologna, Italy}
\centerline{\it I.N.F.N., sezione di Bologna, Italy}
\vskip.9cm
\hrule
\vskip.6cm
\centerline{\bf Abstract} 
\vskip.4cm
\par\noindent
We study a formulation of the standard Poisson sigma model in which
the target space Poisson manifold carries the Hamilton action of some 
finite dimensional Lie algebra.
We show that the structure of the action and the properties of the gauge 
invariant observables can be understood in terms of the associated target 
space equivariant cohomology.
We use a de Rham superfield formalism to efficiently explore the implications 
of the Batalin--Vilkoviski master equation. 
\par\noindent
MSC-class: 53D17, 53B50. Keywords: Poisson Sigma Model, Poisson Geometry,
Cohomology.
\vfill\eject

\vskip .4cm
{\bf 0. Introduction}
\vskip .4cm
\par

A Poisson manifold is a manifold equipped with a Poisson structure.
The Poisson sigma model associates to any Poisson manifold a two 
dimensional sigma model having the Poisson manifold as target space
\ref{1--4}. By means of suitable choices of the Poisson structure, it is 
possible to reproduce a wealth of interesting models, such as two dimensional 
$R^2$ gravity, two dimensional gauge theory and two dimensional 
Wess--Zumino--Witten model. 
More recently, Kontsevich's formulation of deformation quantization 
of the algebra of functions on a Poisson manifold \ref{5} has been interpreted
in terms of the perturbation theory of the corresponding quantum Poisson 
sigma model on the two dimensional disk \ref{6--9}.
This universality and versatility of the Poisson sigma model 
justifies the large body of literature devoted to its study.

In many respects, the Poisson sigma model is a gauge theory 
whose symmetry is based on a Poisson algebra. This makes it interesting, 
but it also poses a number of new problems, especially at the quantum level,
due to the singularity of the kinetic terms and the nonlinearity of the 
symmetry. In this respect, the Batalin--Vilkoviski quantization algorithm 
\ref{10} is essential for achieving a consistent quantization \ref{6,9}.
We feel that a more thorough investigation of the geometry 
of the model is desirable to reach a better handle on these issues.

It is often stated that the Poisson sigma model is a kind of topological
field theory. The known topological field theories are divided in two broad
classes, those of Schwartz type and those of Witten or cohomological type
\ref{11}.
The Poisson sigma model does not seem to fall in either of them.
However, we have found that, when the target space Poisson manifold carries 
the Hamilton action of some finite dimensional Lie algebra, as it happens 
in virtually all the most interesting examples, it has a hidden 
equivariant cohomological structure, that makes it akin to the cohomological 
theories and determines to a considerable degree the structure of the action 
and the properties of the gauge invariant observables. 
The present paper is devoted to the study of this matter.

Our analysis relies to a great extent on an abstract algebraic framework, 
called operation, whose main properties are reviewed in sect. 1. 
Briefly, a $\goth g$ operation over $Z$ consists of a Lie algebra $\goth g$, 
a graded associative algebra $Z$ and a set of derivations $j(\xi)$, $l(\xi)$, 
$\xi\in\goth g$, and $s$ on $Z$ of degrees $-1$, $0$, $+1$, respectively, 
satisfying the graded commutation relations (1.1) below. Every $\goth g$ 
operation over $Z$ admits a canonical $\goth g$ equivariant extension. 
There are three cohomologies associated with the nilpotent 
derivation $s$: ordinary, $\goth g$ basic and $\goth g$ equivariant.

The elements of our construction are provided by the rich geometry of 
Poisson manifolds, whose basic facts are reviewed in sect. 2.
The main geometric datum of a Poisson manifold $M$ is a $2$-vector 
$\varpi^{ij}$, satisfying the Poisson condition (2.1) below, 
in terms of which the Poisson brackets $\{\,,\}$ of $M$ are defined. 
The Lie algebra of Poisson vector fields of $M$ is the symmetry Lie algebra 
of the Poisson structure of $M$ and thus plays an essential role. 
In applications, however, it is often natural to restrict 
oneself to the Lie subalgebra of Hamilton vector fields.

When a Poisson manifold $M$ carries the Poisson or Hamilton action of some 
Lie algebra $\goth h$, one can define a $\goth h$ operation over the space 
of functions of the superbundle $\Pi T\Pi T^*M$ and its $\goth h$ equivariant 
extension. $\Pi T\Pi T^*M$ allows for a natural unified description of the 
induced $\goth h$ action on multivectors and forms of $M$. This construction 
is expounded in detail in sects. 3, 4, 5, 6 and 7. 

Forms on a $2$--dimensional manifold $\Sigma$ can be viewed as elements of 
the space of functions on superbundle $\Pi T\Sigma$, which we shall call 
de Rham superfields. This formalism, which is illustrated in sect. 8, turns 
out to be elegant and convenient.

One can construct a de Rham superfield realization of the $\goth h$ 
equivariant operation over $\Pi T\Pi T^*M$ simply by promoting 
each of its generators $x^i$, $y_i$, $\gamma^a$, etc. to a de Rham 
superfield. This leads to a $\goth h$ operation over a formal graded 
associative algebra of superfields, referred to as $\goth h$ Hamilton de Rham 
superfield operation. The $\goth h$ basic cohomology of this is 
intimately related to the $\goth h$ equivariant cohomology of $\Pi T\Pi T^*M$. 
Although obvious, this fact lies at the heart of our  
analysis of the Poisson sigma model, expounded in sects.  9, 10, and 11, 
which we now outline briefly.

Consider a Poisson manifold $M$ with Poisson $2$--vector $\varpi^{ij}$
carrying the Hamilton action of some finite 
dimensional Lie algebra $\goth h$. Let $h_a$ be the functions of $M$ 
corresponding via the action to fiducial generators $t_a$ of $\goth h$
with structure constants $c^c{}_{ab}$, so that  
$$
\{h_a,h_b\}=c^c{}_{ab}h_c.
\eqno(0.1)
$$
Finally, let $\pi^{ij}$ be a $2$--vector of $M$. The action of the 
Poisson sigma model is 
$$
{\cal S}_\pi=\int_\Sigma\mu\big(y_idx^i+\hbox{$1\over 2$}\pi^{ij}(x)y_iy_j
-d\gamma^ah_a(x)\big),
\eqno(0.2)
$$
where $x^i$, $y_i$, $\gamma^a$ are generators of the $\goth h$ Hamilton de Rham
superfield operation and $\mu$ is the integration supermeasure of $\Sigma$.
${\cal S}_\pi$ satisfies the Batalin--Vilkoviski classical
master equation if $\pi^{ij}$ is a Poisson $2$--vector and if
$$
\pi^{ij}\partial_jh_a=0.
\eqno(0.3)
$$
So, $\pi^{ij}$ defines another Poisson structure on $M$
with respect to which the functions $h_a$ are Casimir. 
However, we stress that, when we refer to $M$ as a Poisson manifold, the 
underlying Poisson structure implied is that associated to $\varpi^{ij}$. 
The crucial result is 
$$
j(r){\cal S}_\pi=0, \quad l(r){\cal S}_\pi=0,\quad s{\cal S}_\pi=0,
\eqno(0.4)
$$
for $r\in\goth h$, if $\pi^{ij}$ Schouten commutes with $\varpi^{ij}$.
Eq. (0.4) then states that ${\cal S}_\pi$ is a representative of a degree $0$ 
$\goth h$ basic cohomology class of Hamilton de Rham superfield operation.
The Batalin--Vilkoviski nilpotent variation $\delta_\pi$ and the derivations 
$j(r)$, $l(r)$ and $s$ (anti)commute. 
Thus, the Batalin--Vilkoviski cohomology and 
the $\goth h$ basic Hamilton de Rham superfield cohomology are compatible.

When $\pi^{ij}=\varpi^{ij}$ the $\goth h$ equivariant cohomology
of $\Pi T\Pi T^*M$ degenerates in the ordinary one, because, by (0.3), 
the Hamilton action is given through Casimir functions and thus is trivial.
Correspondingly, the $\goth h$ basic Hamilton de Rham superfield cohomology 
reduces to the Batalin--Vilkoviski cohomology.

This is not as disappointing as it may seem at first sight.
It often happens that a complicated Poisson $2$--vector $\pi^{ij}$ can be 
written as 
$$
\pi^{ij}=\varpi^{ij}+\vartheta^{ij},
\eqno(0.5)
$$
where $\varpi^{ij}$ is a well understood Poisson $2$--vector and 
$\vartheta^{ij}$ is another Poisson $2$--vector Schouten commuting with 
$\varpi^{ij}$. One can then view $\pi^{ij}$ as a perturbation of 
$\varpi^{ij}$ and try to understand the $\pi^{ij}$ Poisson sigma model
as a perturbation of the $\varpi^{ij}$ Poisson sigma model.
There are plenty of such examples, such as the $R^2$ gravity Poisson sigma 
model and the affine Lie--Poisson sigma model. More examples are illustrated 
in sect. 12.

All known topological field theories of cohomological 
type are characterized by a Lie group $\cal G$, a space of fields $\cal P$
carrying a right $\cal G$ action, a certain $\cal G$ operation over $\cal P$ 
and an action $S$ that is a representative of a degree $0$ $\cal G$ basic or 
equivariant cohomology class of $\cal P$ \ref{12}. Apparently, the Poisson 
sigma model has a very similar structure. The analogy, however, stops 
here. In the topological field theories of cohomological type, the group 
$\cal G$ is infinite dimensional and the action $S$ is a Mathai-Quillen 
representative of the Thom class of some vector bundle $\cal E$ over 
${\cal P}/{\cal G}$ and describes localization on the zero locus of certain 
sections of $\cal E$ \ref{13--14}. So far, a similar interpretation does not 
seem to be possible for the version of the Poisson sigma model studied 
in this paper.

\eject
\vskip .4cm
{\bf 1. Generalities on Basic and Equivariant Cohomology}
\vskip .4cm
\par
Our analysis of the Poisson sigma model is based on a formal algebraic 
framework, called operation, whose main properties we shall now review
(see ref. \ref{15} for background material). 

A $\goth g$ operation over $Z$ is a quintuplet
$(Z,\goth g,j,l,s)$, where $\goth g$ is a Lie algebra, $Z$ is a graded 
associative algebra and $j(\xi)$, $l(\xi)$, $\xi\in\goth g$, and $s$ are 
graded derivations on $Z$ of degree $-1$, $0$, $+1$, respectively, satisfying 
Cartan's algebra:
$$
\eqalignno{&\hbox to 2.05truecm{$[j(\xi),j(\eta)]$\hfill}
\hbox to 3.5truecm{$=0$,\hfill}
\hbox to 2.03truecm{$[l(\xi),j(\eta)]$\hfill}
\hbox to 3.5truecm{$=j([\xi,\eta])$,\hfill}\vphantom{\Big[}&(1.1)\cr
&\hbox to 2.05truecm{$[l(\xi),l(\eta)]$\hfill}
\hbox to 3.5truecm{$=l([\xi,\eta])$,\hfill}
\hbox to 2.03truecm{$[s,j(\eta)]$\hfill}
\hbox to 3.5truecm{$=l(\eta)$,\hfill}\vphantom{\Big[}&\cr
&\hbox to 2.05truecm{$[s,l(\eta)]$\hfill}
\hbox to 3.5truecm{$=0$,\hfill}
\hbox to 2.03truecm{$[s,s]$\hfill}
\hbox to 3.5truecm{$=0$,\hfill}\vphantom{\Big[}&\cr}
$$
where the above are graded commutators. In all examples considered 
below, the graded algebra $Z$ is finitely generated. So, the graded 
derivations $j(\xi)$, $l(\xi)$ and $s$ are completely defined by their action 
on a suitable set of natural generators.

Since $s^2=0$, one can 
define the cohomology of the differential complex $(Z,s)$. This is called 
ordinary cohomology of $Z$. More importantly, one may consider 
the differential complex $(Z_{\rm basic},s)$, where $Z_{\rm basic}$ is the 
$s$ invariant subalgebra of $Z$ annihilated by all $j(\xi)$, $l(\xi)$, 
$\xi\in\goth g$. The corresponding cohomology is referred to
as $\goth g$ basic cohomology of $Z$.

To any Lie algebra $\goth g$, there is canonically associated the Weil 
operation $(W(\goth g),\goth g,j,l,$ $s)$. 
Here, $W(\goth g)=A(\goth g^\vee)\otimes S(\goth g^\vee)$, 
where $A(\goth g^\vee)$, $S(\goth g^\vee)$ are the antisymmetric, symmetric 
algebras of $\goth g^\vee$, respectively. The natural $\goth g$--valued
generators $\omega$, $\Omega$ of $W(\goth g)$ carry degrees $1$, $2$, 
respectively, and satisfy
$$
\eqalignno{&\hbox to 1.12truecm{$j(\xi)\omega$\hfill}
\hbox to 4.0truecm{$=\xi$,\hfill}
\hbox to 1.12truecm{$j(\xi)\Omega$\hfill}
\hbox to 4.0truecm{$=0$,\hfill}\vphantom{\Big[}&(1.2)\cr
&\hbox to 1.12truecm{$l(\xi)\omega$\hfill}
\hbox to 4.0truecm{$=-[\xi,\omega]$,\hfill}
\hbox to 1.12truecm{$l(\xi)\Omega$\hfill}
\hbox to 4.0truecm{$=-[\xi,\Omega]$,\hfill}\vphantom{\Big[}&\cr
&\hbox to 1.12truecm{$s\omega$\hfill}
\hbox to 4.0truecm{$=\Omega-(1/2)[\omega,\omega]$,\hfill}
\hbox to 1.12truecm{$s\Omega$\hfill}
\hbox to 4.0truecm{$=-[\omega,\Omega]$.\hfill}\vphantom{\Big[}&\cr}
$$
The cohomology of $W(\goth g)$ is trivial. The $\goth g$ basic cohomology 
of $W(\goth g)$ is isomorphic to the $\ad\goth g$ invariant subalgebra
of $S(\goth g^\vee)$, 

For a given operation $(Z,\goth g,j,l,s)$, one can construct the operation 
$(Z\hat\otimes W(\goth g),\goth g,j,l,s)$, where $\hat\otimes$ denotes
graded tensor product. $(Z\hat\otimes W(\goth g),\goth g,j,l,s)$ is called 
equivariant extension of $(Z,\goth g,j,l,s)$. The $\goth g$ basic cohomology 
of $Z\hat\otimes W(\goth g)$ is called $\goth g$ equivariant cohomology of 
$Z$. Representatives of $\goth g$ equivariant cohomology classes of $Z$ in 
$Z\hat\otimes W(\goth g)$ yield representatives of $\goth g$ 
basic cohomology classes of $Z$ by replacing the 
Weil generators $\omega$, $\Omega$, by a connection $a$, $A$ of $Z$, 
i. e. a pair of elements $a$, $A$ of $Z\otimes\goth g$ of degree $1$, $2$, 
respectively, satisfying (1.2) with $\omega$, $\Omega$ replaced by $a$, $A$ 
(Weil homomorphism).  

A homomorphism of the operation $(Z,\goth g,j,l,s)$ into the operation 
$(Z',\goth g',j',l',s')$ is a pair $(f,\mu)$, where $f:Z\rightarrow Z'$ 
is a degree zero graded algebra homomorphism, 
$\mu:\goth g'\rightarrow \goth g$ is a Lie algebra homomorphism and 
$$
j'(\xi')\circ f= f\circ j(\mu(\xi')), \quad
l'(\xi')\circ f= f\circ l(\mu(\xi')), \quad
s'\circ f= f\circ s,
\eqno(1.3)
$$
with $\xi'\in\goth g'$. Clearly, $(f,\mu)$ induces a homomorphism of the
$\goth g$ basic cohomology of $Z$ into the $\goth g'$ 
basic cohomology of $Z'$.

If $\mu:\goth g'\rightarrow \goth g$ is a Lie algebra homomorphism, 
then the pair $(\mu^\vee,\mu)$ is a homomorphism of the Weil 
operation of $\goth g$ $(W(\goth g),\goth g,j,l,s)$ into
the Weil operation of $\goth g'$ $(W(\goth g'),\goth g',j',l',s')$,
where $\mu^\vee$ denotes the natural extension to $W(\goth g)$ of the
dual linear homomorphism $\mu^\vee:\goth g^\vee\rightarrow \goth g'^\vee$.
The generators $\omega$, $\Omega$, $\omega'$, $\Omega'$ of 
$W(\goth g)$, $W(\goth g')$, respectively, obey the important relations
$$
\mu^\vee(\omega)=\mu(\omega'),\quad \mu^\vee(\Omega)=\mu(\Omega').
\eqno(1.4)
$$
By combining these identities with (1.3) with $f=\mu^\vee$, we find that
the action of $j'(\xi')$, $l'(\xi')$. $\xi'\in \goth g'$ and $s'$
on $\mu(\omega')$, $\mu(\Omega')$ is obtained from that of  
$j(\mu(\xi'))$, $l(\mu(\xi'))$ and $s$ on $\omega$, $\Omega$, eq. 
(1.2), by applying $\mu^\vee$.

If $(f,\mu)$ is a homomorphism of the operations $(Z,\goth g,j,l,s)$, 
$(Z',\goth g',j',l',s')$, $(f\hat\otimes\mu^\vee,\mu)$ is a 
homomorphism of the corresponding equivariant extensions
$(Z\hat\otimes W(\goth g),\goth g,j,l,s)$, $(Z'$ $\hat\otimes W(\goth g'),$
$\goth g',j',l',s')$ and induces a homomorphism of the $\goth g$ 
equivariant cohomology of $Z$ into the $\goth g'$ equivariant cohomology 
of $Z'$.

If $(Z,\goth g,j,l,s)$ is an operation and $\mu:\goth g'\rightarrow \goth g$ 
is a Lie algebra homomorphism, we can define a new operation
$(Z,\goth g',j',l',s')$ by setting
$$
j'(\xi')=j(\mu(\xi')), \quad
l'(\xi')=l(\mu(\xi')), \quad
s'=s,
\eqno(1.5)
$$
with $\xi'\in\goth g$. $(Z,\goth g',j',l',s')$ is called pull--back of 
$(Z,\goth g,j,l,s)$ by $\mu$. In the particular case where $\goth g'$ is a 
Lie subalgebra of $\goth g$ and $\mu$ is the natural inclusion, 
$(Z,\goth g',j',l',s')$ is called restriction of $(Z,\goth g,j,l,s)$ to 
$\goth g'$. $(\id_Z,\mu)$ is a homomorphism of the
operations $(Z,\goth g,j,l,s)$, $(Z,\goth g',j',l',s')$, which induces
an injection of the $\goth g$ basic cohomology into the $\goth g'$ 
basic cohomology of $Z$.
Similarly, $(\id_Z\hat\otimes \mu^\vee,\mu)$ is a homomorphism of the
associated equivariant extensions $(Z\hat\otimes W(\goth g),\goth g,j,l,s)$, 
$(Z\hat\otimes W(\goth g'),\goth g',j',l',s')$, which induces
a homomorphism of the $\goth g$ equivariant cohomology 
into the $\goth g'$ equivariant cohomology of $Z$.

A wide class of operations is built as follows. Consider:
$i$) a Lie group $\cal G$ with Lie algebra $\Lie{\cal G}$;
$ii$) a principal $\cal G$ bundle $\pi_{\cal P}:\cal P\to\cal M$.
Define $Z=\Omega^*({\cal P})$, $\goth g=\Lie{\cal G}$,  
$j(\xi)=j_{\cal P}(C_\xi)$, $l(\xi)=l_{\cal P}(C_\xi)$, for $\xi\in\goth g$, 
and $s=d_{\cal P}$, where $\Omega^*({\cal P})$ is the graded algebra of 
differential forms of $\cal P$, $j_{\cal P}$, $l_{\cal P}$ and $d_{\cal P}$
are the customary differential geometric contraction, Lie derivative and 
de Rham differential operators and $C_\xi$ is the vertical vector field 
corresponding to $\xi$. The resulting quintuplet $(Z,\goth g,j,l,s)$ 
is an operation. It basic cohomology is isomorphic to the de Rham cohomology 
of the base $\cal M$.

Typically, in cohomological topological quantum field theory, $\cal G$
is a gauge group, $\cal P$ is a supermanifold of gauge and matter fields 
propagating on a space--time manifold $X$ and $\cal M$ is some sort of gauge 
orbit or moduli space \ref{11}. 
Representatives of equivariant classes, known as topological 
observables, play an important role, since they yield via the Weil 
homomorphism forms on $\cal M$, which can be used to probe its structure.
They are obtained by integrating on cycles of $X$ certain differential 
forms of $X$ built with the fields.  One is thus interested in mod $d$ 
equivariant classes, where $d$ is the Rham differential of $X$. 
This leads to the well-known descent formalism \ref{11}.

\vskip .4cm
{\bf 2. Generalities on Poisson Manifolds}
\vskip .4cm
\par
The target space of the Poisson sigma model is a Poisson manifold. We 
shall now briefly review the properties of Poisson manifolds relevant in the 
following discussion (see ref. \ref{16} for background material). 

By definition, a manifold $M$ is Poisson if it is equipped with a $2$--vector
$\varpi^{ij}$ satisfying the relation
$$
[\varpi,\varpi]=0~~\hbox{i. e.}~~
\varpi^{il}\partial_l\varpi^{jk}+\varpi^{jl}\partial_l\varpi^{ki}
+\varpi^{kl}\partial_l\varpi^{ij}=0.
\eqno(2.1)
$$
Here and below, $[\cdot,\cdot]$ denotes the Schouten brackets, the only 
natural pairing of multivectors defined on any manifold.

On a Poisson manifold, one can define Poisson brackets of (local)
functions $f$, $g$ 
$$
\{f,g\}=\varpi^{ij}\partial_if\partial_jg.
\eqno(2.2)
$$
On account of (2.1), one indeed has 
$$
\{f,g\}+\{g,f\}=0,
\eqno(2.3)
$$
$$
\{f,\{g, h\}\}+\{g,\{h, f\}\}+\{h,\{f,g\}\}=0.
\eqno(2.4)
$$
In this way, the space of functions $\Fun(M)$ becomes a Lie algebra,
the Poisson algebra of $M$.

Functions $f$ satisfying 
$$
[\varpi,f]=0~~\hbox{i. e.}~~\varpi^{ij}\partial_jf=0
\eqno(2.5)
$$
are called Casimir functions. Such $f$ Poisson commute with any function 
$g$: $\{f,g\}=0$. So, they form the center $\Cas(M)$ 
of $\Fun(M)$. 

Vector fields $u^i$ such that
$$
[\varpi,u]=0~~\hbox{i. e.}~~u^k\partial_k\varpi^{ij}-\partial_ku^i\varpi^{kj}
-\partial_ku^j\varpi^{ik}=0
\eqno(2.6)
$$
are called Poisson vector fields. Such $u^i$ are precisely the vector 
fields leaving $\varpi^{ij}$ invariant, since (2.6) states that
$l_M(u)\varpi^{ij}=0$, $l_M(u)$ being the usual Lie derivative.
The Poisson vector fields $u^i$ of the form
$$
u_f=-[\varpi,f]~~\hbox{i. e.}~~u_f{}^i=-\varpi^{ij}\partial_jf
\eqno(2.7)
$$
for some function $f$ are called Hamilton vector fields.

The Poisson vector fields span a Lie subalgebra $\Pois(M)$ 
of the Lie algebra $\Vect(M)$ of the vector fields of $M$.
The Poisson vector fields are the natural symmetry Lie algebra of a Poisson 
manifold and of the associated geometrical structures.
The Hamilton vector fields form a Lie subalgebra $\Ham(M)$
of $\Pois(M)$ as
$$
[u_f,u_g]=u_{\{f,g\}},
\eqno(2.8)
$$
for any two functions $f$, $g$. So, (2.7) establishes a canonical surjective 
Lie algebra homomorphism $\Fun(M)\rightarrow\Ham(M)$ with
kernel $\Cas(M)$.

On a Poisson manifold, one can define a natural degree $+1$ operator acting 
on a general $p$--vector $\zeta^{i_1\cdots i_p}$
$$
q\zeta=[\varpi,\zeta].
\eqno(2.9)
$$
Using (2.1), it is easy to verify that
$$
q{}^2=0.
\eqno(2.10)
$$
Hence, a Poisson manifold has a natural notion of cohomology, called
Poisson--Lichnero\-wicz  cohomology. Since 
$$
[q,l_M(u)]=0,
\eqno(2.11)
$$
for any Poisson vector field $u^i$, one can define a Poisson (Hamilton)
invariant Poisson--Lichnero\-wicz cohomology by restricting $q$ to the 
complex of multivectors $\zeta^{i_1\cdots i_p}$ such that 
$l_M(u)\zeta^{i_1\cdots i_p}=0$ for all Poisson vector fields $u^i$
($l_M(u_f)\zeta^{i_1\cdots i_p}=0$ for all functions $f$). 

The $0$ Poisson--Lichnerowicz cocycles are the Casimir functions and span the 
$0$-th Poisson--Lichnerowicz cohomology. The  $1$ Poisson--Lichnerowicz 
cocycles are the Poisson vector fields, the $1$ Poisson--Lichnerowicz 
coboundaries are the Hamilton vector fields. So, the $1$--st 
Poisson--Lichnerowicz cohomology is the quotient of the Poisson by the 
Hamilton vector field spaces. 
It is possible to show that the $2$--st Poisson--Lichnerowicz cohomology
describes the space of the infinitesimal deformation $\beta^{ij}$
of the Poisson $2$--vector $\varpi^{ij}$ modulo the deformations of the form
$l_M(u)\varpi^{ij}$ for some vector field $u^i$.

\vskip .4cm
{\bf 3. The Poisson and the Poisson equivariant Operation of $\Pi TM$}
\vskip .4cm
\par

We are now going to construct the operations relevant in the following 
analysis. For a given Poisson manifold $M$, all these are $\goth H$ 
operations over the graded associative algebra $\Fun(\Pi T\Pi T^*M)$ of 
functions of the superbundle $\Pi T\Pi T^*M$, or its equivariant extension 
$\Fun(\Pi T\Pi T^*M)\hat\otimes W(\goth H)$, where $\goth H$ is some 
relevant Lie subalgebra of $\Vect(M)$ and $\Pi$ denotes the fiber parity 
inversion functor. Further, they all have a natural set of generators 
$x^i$, $X^\sharp{}^i$, $y_i$, $Y^\sharp{}_i$ with the following degree 
assignments
$$
\eqalignno{
&\hbox to 2.5truecm{$\deg x^i=0$,\hfill}
\hbox to 2.5truecm{$\deg X^\sharp{}^i=1$,\hfill}&(3.1)\cr 
&\hbox to 2.5truecm{$\,\deg y_i=1$,\hfill}
\hbox to 2.5truecm{$\,\deg Y^\sharp{}_i=2$,\hfill}&\cr}
$$
and transformation properties under a change of local
coordinates $t\rightarrow t'$
$$
\eqalignno{&\hbox to 3.5truecm{$x'^i=t'^i(x)$,\hfill}
\hbox to 8.0truecm{$X^\sharp{}'^i
={\partial t'^i \over\partial t^j}(x) X^\sharp{}^j$,~~~~\hfill}
\vphantom{\Big [}&(3.2)\cr 
&\hbox to 3.5truecm{$\,\,y'_i={\partial t^j \over\partial t'^i}(x)y_j$,\hfill}
\hbox to 8.0truecm{$\,\,Y^\sharp{}'_i={\partial t^j \over\partial t'^i}(x) 
Y^\sharp{}_j+\Big({\partial^2 t^j \over\partial t'^i\partial t'^k}
{\partial t'^k \over\partial t^l}\Big)(x) X^\sharp{}^ly_j$.\hfill}
\vphantom{\Big [}&\cr}
$$

There are canonical injections of the spaces of $p$--vectors and $p$--forms 
of $M$ into the graded algebra $\Fun(\Pi T\Pi T^*M)$ or its equivariant 
extension $\Fun(\Pi T\Pi T^*M)\hat\otimes W(\goth H)$. 
To any $p$--vector $\beta^{i_1\cdots i_p}$, one associates 
$$
\beta(x,y)=\hbox{$1\over p!$}\beta^{i_1\cdots i_p}(x)y_{i_1}\cdots y_{i_p}.
\vphantom{\bigg[}
\eqno(3.3)
$$
Similarly, to any $p$--form $\sigma_{i_1\cdots i_p}$, one associates 
$$
\sigma(x, X^\sharp)=\hbox{$1\over p!$}\sigma_{i_1\cdots i_p}(x)
 X^\sharp{}^{i_1}\cdots  X^\sharp{}^{i_p}.
\vphantom{\bigg[}
\eqno(3.4)
$$
These representations are very convenient and will be used throughout.

The fundamental differential operation of $\Pi TM$ is defined independently 
from any differential geometric structure on $M$. It is the $\Vect(M)$ 
operation over $\Fun(\Pi T\Pi T^*M)$ with natural generators $x^i$, $ X^i$, 
$y_i$, $ Y_i$ obeying
$$
\eqalignno{&\hbox to 1.4truecm{$j(u)x^i$\hfill}
\hbox{$=0$\hfill},\vphantom{\Big[}&(3.5)\cr 
&\hbox to 1.4truecm{$l(u)x^i$\hfill}
\hbox{$=u^i(x)$\hfill},\vphantom{\Big[}&\cr 
&\hbox to 1.4truecm{$sx^i$\hfill}
\hbox{$= X^i$\hfill},\vphantom{\Big[}&\cr 
&\hbox to 1.4truecm{$j(u) X^i$\hfill}
\hbox{$=u^i(x)$\hfill},\vphantom{\Big[}&\cr 
&\hbox to 1.4truecm{$l(u) X^i$\hfill}
\hbox{$=\partial_ju^i(x) X^j$\hfill},\vphantom{\Big[}&\cr 
&\hbox to 1.4truecm{$s X^i$\hfill}
\hbox{$=0$\hfill},\vphantom{\Big[}&\cr 
&\hbox to 1.4truecm{$j(u)y_i$\hfill}
\hbox{$=0$\hfill},\vphantom{\Big[}&\cr
&\hbox to 1.4truecm{$l(u)y_i$\hfill}
\hbox{$=-\partial_iu^j(x)y_j$\hfill},\vphantom{\Big[}&\cr 
&\hbox to 1.4truecm{$sy_i$\hfill}
\hbox{$= Y_i$\hfill},\vphantom{\Big[}&\cr
&\hbox to 1.4truecm{$j(u) Y_i$\hfill}
\hbox{$=-\partial_iu^j(x)y_j$\hfill},\vphantom{\Big[}&\cr
&\hbox to 1.4truecm{$l(u) Y_i$\hfill}
\hbox{$=-\partial_iu^j(x) Y_j
-\partial_i\partial_ju^k(x) X^jy_k$\hfill},\vphantom{\Big[}&\cr 
&\hbox to 1.4truecm{$s Y_i$\hfill}
\hbox{$=0$\hfill},\vphantom{\Big[}&\cr} 
$$
with $u^i$ any vector field in $\Vect(M)\vphantom{\Big[}$.

If $M$ is a Poisson manifold with Poisson $2$--vector $\varpi^{ij}$, 
the natural symmetry Lie algebra is 
the subalgebra of $\Vect(M)$ leaving $\varpi^{ij}$ invariant, that is 
the Poisson Lie algebra 
$\Pois(M)$ (cfr. sect. 2). The Poisson differential operation of $\Pi TM$
is obtained by restricting the fundamental differential operation of $\Pi TM$
to $\Pois(M)$ (cfr. sect. 1). It is the $\Pois(M)$ operation over 
$\Fun(\Pi T\Pi T^*M)$ with natural generators $x^i$, $X^*{}^i$, 
$y_i$, $Y^*{}_i$ obeying
$$
\eqalignno{&\hbox to 1.63truecm{$j(u)x^i$\hfill}
\hbox{$=0$\hfill},\vphantom{\Big[}&(3.6)\cr
&\hbox to 1.63truecm{$l(u)x^i$\hfill}
\hbox{$=u^i(x)$\hfill},\vphantom{\Big[}&\cr
&\hbox to 1.63truecm{$sx^i$\hfill}
\hbox{$= X^{*i}+\varpi^{ij}(x)y_j$\hfill},\vphantom{\Big[}&\cr
&\hbox to 1.63truecm{$j(u) X^*{}^i$\hfill}
\hbox{$=u^i(x)$\hfill},\vphantom{\Big[}&\cr
&\hbox to 1.63truecm{$l(u) X^*{}^i$\hfill}
\hbox{$=\partial_ju^i(x) X^*{}^j$\hfill},\vphantom{\Big[}&\cr
&\hbox to 1.63truecm{$s X^*{}^i$\hfill}
\hbox{$=-\varpi^{ij}(x) Y^*{}_j
-\partial_j\varpi^{ik}(x) X^*{}^jy_k$\hfill},\vphantom{\Big[}&\cr
&\hbox to 1.63truecm{$j(u)y_i$\hfill}
\hbox{$=0$\hfill},\vphantom{\Big[}&\cr
&\hbox to 1.63truecm{$l(u)y_i$\hfill}
\hbox{$=-\partial_iu^j(x)y_j$\hfill},\vphantom{\Big[}&\cr
&\hbox to 1.63truecm{$sy_i$\hfill}
\hbox{$= Y^*{}_i+{1\over 2}\partial_i\varpi^{jk}(x)y_jy_k$\hfill},
\vphantom{\Big[}&\cr
&\hbox to 1.63truecm{$j(u) Y^*{}_i$\hfill}
\hbox{$=-\partial_iu^j(x)y_j$\hfill},\vphantom{\Big[}&\cr
&\hbox to 1.63truecm{$l(u) Y^*{}_i$\hfill}
\hbox{$=-\partial_iu^j(x) Y^*{}_j
-\partial_i\partial_ju^k(x) X^*{}^jy_k$\hfill},\vphantom{\Big[}&\cr
&\hbox to 1.63truecm{$s Y^*{}_i$\hfill}
\hbox{$=-{1\over 2}\partial_i\partial_j\varpi^{kl}(x)
 X^*{}^jy_ky_l+\partial_i\varpi^{jk}(x)y_j Y^*{}_k$\hfill},
\vphantom{\Big[}&\cr}
$$
where now $u^i$ is any Poisson vector field in $\Pois(M)\vphantom{\Big[}$
(cfr. eq. (2.6)). 
The Poisson generators $X^*{}^i$, $Y^*{}_i$ are related to the fundamental 
generators $X^i$, $Y_i$ by the simple shifts 
$$
\eqalignno{&\hbox to 0.83truecm{$X^{*i}$\hfill}
\hbox{$=X^i-\varpi^{ij}(x)y_j$\hfill},\vphantom{\Bigg[}&(3.7)\cr
&\hbox to 0.83truecm{$Y^*{}_i$\hfill}
\hbox{$=Y_i-{1\over 2}\partial_i\varpi^{jk}(x)y_jy_k$\hfill}.
\vphantom{\Big[}&\cr}
$$
Such shifts have a simple geometrical interpretation: 
$$
\eqalignno{&\hbox to 1.0truecm{$q x^i$\hfill}
\hbox{$=-\varpi^{ij}(x)y_j$\hfill},\vphantom{\Big[}&(3.8)\cr
&\hbox to 1.0truecm{$q y_i$\hfill}
\hbox{$=-{1\over 2}\partial_i\varpi^{jk}(x)y_jy_k$\hfill},\vphantom{\Big[}&\cr}
$$
where $q$ is defined in (2.9). 

Next, we consider the Weil operation of the Lie algebra $\Pois(M)$ 
(cfr. sect. 1). It is the $\Pois(M)$ operation over the Weil algebra
$W(\Pois(M))$ with natural Poisson vector field valued generators $\omega^i$, 
$\Omega^i$ satisfying the form of the Weil operation relations (1.2)
appropriate for $\Pois(M)$. It turns out to be more efficient 
to combine $\omega^i$, $\Omega^i$ with the generators $x^i$ to form 
composites $\omega^i(x)$, $\Omega^i(x)$, by replacing the dummy local 
coordinates $t^i$ appearing in the local expression $\omega^i(t)$, 
$\Omega^i(t)$ of $\omega^i$, $\Omega^i$ by $x^i$. 
The Poisson--Weil generators $\omega^i(x)$, $\Omega^i(x)$ carry degrees
$$
\deg\omega^i(x)=1, \qquad \deg\Omega^i(x)=2,\vphantom{\Big[}
\eqno(3.9)
$$
by being Poisson vector fields satisfy 
$$
\eqalignno{&(\omega^k\partial_k\varpi^{ij}-\partial_k\omega^i\varpi^{kj}
-\partial_k\omega^j\varpi^{ik})(x)=0,\vphantom{\Big[}&(3.10)\cr
&(\Omega^k\partial_k\varpi^{ij}-\partial_k\Omega^i\varpi^{kj}
-\partial_k\Omega^j\varpi^{ik})(x)=0,\vphantom{\Big[}&\cr}
$$
and obey
$$
\eqalignno{&\hbox to 1.9truecm{$j(u)\omega^i(x)$\hfill}
\hbox{$=u^i(x)$\hfill},\vphantom{\Big[}&(3.11)\cr
&\hbox to 1.9truecm{$l(u)\omega^i(x)$\hfill}
\hbox{$=\omega^j\partial_ju^i(x)$\hfill},\vphantom{\Big[}&\cr
&\hbox to 1.9truecm{$s\omega^i(x)$\hfill}
\hbox{$=\Omega^i(x)-\omega^j\partial_j\omega^i(x)
+( X^*{}^j+\varpi^{jk}(x)y_k)\partial_j\omega^i(x)$\hfill},
\vphantom{\Big[}&\cr
&\hbox to 1.9truecm{$j(u)\Omega^i(x)$\hfill}
\hbox{$=0$\hfill},\vphantom{\Big[}&\cr
&\hbox to 1.9truecm{$l(u)\Omega^i(x)$\hfill}
\hbox{$=\Omega^j\partial_ju^i(x)$\hfill},\vphantom{\Big[}&\cr
&\hbox to 1.9truecm{$s\Omega^i(x)$\hfill}
\hbox{$=-\omega^j\partial_j\Omega^i(x)+\Omega^j\partial_j\omega^i(x)
+( X^*{}^j+\varpi^{jk}(x)y_k)\partial_j\Omega^i(x)$\hfill},
\vphantom{\Big[}&\cr}
$$
where $u^i$ is any Poisson vector field in $\Pois(M)$.

The equivariant extension of the Poisson differential operation 
of $\Pi TM$ (cfr. sect. 1), referred to as Poisson equivariant differential 
operation in the following, is now easily worked out. It is the 
$\Pois(M)$ operation over $\Fun(\Pi T\Pi T^*M)\hat\otimes W(\Pois(M))$ 
with natural generators $x^i$, $\tilde X^i$, $y_i$, $\tilde Y_i$, 
$\omega^i(x)$, $\Omega^i(x)$ satisfying 
$$
\eqalignno{&\hbox to 1.85truecm{$j(u)x^i$\hfill}
\hbox{$=0$\hfill},\vphantom{\Big[}&(3.12)\cr
&\hbox to 1.85truecm{$l(u)x^i$\hfill}
\hbox{$=u^i(x)$\hfill},\vphantom{\Big[}&\cr
&\hbox to 1.85truecm{$sx^i$\hfill}
\hbox{$=\tilde X^i+\varpi^{ij}(x)y_j+\omega^i(x)$\hfill},\vphantom{\Big[}&\cr
&\hbox to 1.85truecm{$j(u)\tilde X^i $\hfill}
\hbox{$=0$\hfill},\vphantom{\Big[}&\cr
&\hbox to 1.85truecm{$l(u)\tilde X^i $\hfill}
\hbox{$=\partial_ju^i(x)\tilde X^j$\hfill},\vphantom{\Big[}&\cr
&\hbox to 1.85truecm{$s\tilde X^i$\hfill}
\hbox{$=-\varpi^{ij}(x)\tilde Y_j
-\partial_j\varpi^{ik}(x)\tilde X^jy_k-\Omega^i(x)
+\partial_j\omega^i(x)\tilde X^j$\hfill},\vphantom{\Big[}&\cr
&\hbox to 1.85truecm{$j(u)y_i$\hfill}
\hbox{$=0$\hfill},\vphantom{\Big[}&\cr
&\hbox to 1.85truecm{$l(u)y_i$\hfill}
\hbox{$=-\partial_iu^j(x)y_j$\hfill},\vphantom{\Big[}&\cr
&\hbox to 1.85truecm{$sy_i$\hfill}
\hbox{$=\tilde Y_i+{1\over 2}\partial_i\varpi^{jk}(x)y_jy_k
-\partial_i\omega^j(x)y_j$\hfill},\vphantom{\Big[}&\cr
&\hbox to 1.85truecm{$j(u)\tilde Y_i$\hfill}
\hbox{$=0$\hfill},\vphantom{\Big[}&\cr
&\hbox to 1.85truecm{$l(u)\tilde Y_i$\hfill}
\hbox{$=-\partial_iu^j(x)\tilde Y_j
-\partial_i\partial_ju^k(x)\tilde X^jy_k$\hfill},\vphantom{\Big[}&\cr
&\hbox to 1.85truecm{$s\tilde Y_i$\hfill}
\hbox{$=-{1\over 2}\partial_i\partial_j\varpi^{kl}(x)
\tilde X^jy_ky_l+\partial_i\varpi^{jk}(x)y_j\tilde Y_k
+\partial_i\Omega^j(x)y_j$\hfill}\vphantom{\Big[}&\cr
&\hbox to 1.85truecm{$\hphantom{s\tilde Y_i}$\hfill}
\hbox{$\hphantom{=}-\partial_i\omega^j(x)\tilde Y_j
-\partial_i\partial_j\omega^k(x)\tilde X^jy_k$\hfill},\vphantom{\Big[}&\cr
&\hbox to 1.85truecm{$j(u)\omega^i(x)$\hfill}
\hbox{$=u^i(x)$\hfill},\vphantom{\Big[}&(3.13)\cr
&\hbox to 1.85truecm{$l(u)\omega^i(x)$\hfill}
\hbox{$=\omega^j\partial_ju^i(x)$\hfill},\vphantom{\Big[}&\cr
&\hbox to 1.85truecm{$s\omega^i(x)$\hfill}
\hbox{$=\Omega^i(x)
+(\tilde X^j+\varpi^{jk}(x)y_k)\partial_j\omega^i(x)$\hfill},
\vphantom{\Big[}&\cr
&\hbox to 1.85truecm{$j(u)\Omega^i(x)$\hfill}
\hbox{$=0$\hfill},\vphantom{\Big[}&\cr
&\hbox to 1.85truecm{$l(u)\Omega^i(x)$\hfill}
\hbox{$=\Omega^j\partial_ju^i(x)$\hfill},\vphantom{\Big[}&\cr
&\hbox to 1.85truecm{$s\Omega^i(x)$\hfill}
\hbox{$=\Omega^j\partial_j\omega^i(x)
+(\tilde X^j+\varpi^{jk}(x)y_k)\partial_j\Omega^i(x),$\hfill}
\vphantom{\Big[}&\cr}
$$
where again $u^i$ is any Poisson vector field in $\Pois(M)\vphantom{\Big[}$.
The Poisson equivariant generators $\tilde X^i$, $\tilde Y_i$
are related to the Poisson generators $X^*{}^i$, $Y^*{}_i$ 
by the simple shifts 
$$
\eqalignno{&\hbox to 0.65truecm{$\tilde X^i$\hfill}
\hbox{$= X^*{}^i-\omega^i(x)$\hfill},\vphantom{\bigg[}&(3.14)\cr
&\hbox to 0.65truecm{$\tilde Y_i$\hfill}
\hbox{$= Y^*{}_i+\partial_i\omega^j(x)y_j$\hfill}.\vphantom{\Big[}&\cr}
$$
The shifts have a simple formal geometrical interpretation: 
$$
\eqalignno{&\hbox to 1.2truecm{$l(\omega)x^i$\hfill}
\hbox{$=\omega^i(x)$\hfill},\vphantom{\Big[} &(3.15)\cr
&\hbox to 1.2truecm{$l(\omega)y_i$\hfill}
\hbox{$=-\partial_i\omega^j(x)y_j$\hfill}.\vphantom{\Big[}&\cr}
$$
Note that all the Poisson equivariant generators but $\omega^i(x)$
are horizontal, i. e. they are annihilated by
every $j(u)$ for all Poisson vector fields $u^i$ in $\Pois(M)$. 

\vskip .4cm
{\bf 4. The Hamilton and the Hamilton equivariant Operation of $\Pi TM$}
\vskip .4cm
\par
As discussed in sect. 2, if $M$ is a Poisson manifold, 
there is a canonical Lie algebra homomorphism
$\varrho:\Fun(M)\rightarrow\Pois(M)$, defined by (2.7), of the Poisson algebra
$\Fun(M)$ into the Poisson vector field Lie algebra $\Pois(M)$, whose image is
the Hamilton vector field Lie subalgebra $\Ham(M)$.

The Hamilton differential operation of $\Pi TM$ 
is the pull--back of the Poisson differential operation of $\Pi TM$ 
by the Lie algebra homomorphism $\varrho$ (cfr. sect. 1). 
It is the $\Fun(M)$ operation over $\Fun(\Pi T\Pi T^*M)$
with natural generators $x^i$, $X^*{}^i$, $y_i$, $Y^*{}_i$
satisfying (3.6) with with $j(u)$, $l(u)$, $u^i$ substituted by 
$j(f)$, $l(f)$, $u_f{}^i$ (cfr. eq. (2.7)), respectively, for any function 
$f$ of $\Fun(M)$. The Hamilton generators $X^*{}^i$, $Y^*{}_i$ are still 
related to the fundamental generators $X^i$, $Y_i$ by $(3.7)$. 

Next, we consider the Weil operation of the Lie algebra $\Fun(M)$ 
(cfr. sect. 1). It is the $\Fun(M)$ operation over the Weil algebra
$W(\Fun(M))$ with natural scalar valued generators $\phi$, 
$\Phi$ satisfying the form of the Weil operation relations (1.2)
appropriate for $\Fun(M)$. As in the Poisson case, it turns out to be more 
efficient to combine $\phi$, $\Phi$ with the generators $x^i$ to form 
composites $\phi(x)$, $\Phi(x)$. The Hamilton--Weil generators $\phi(x)$, 
$\Phi(x)$ carry degrees 
$$
\deg\phi(x)= 1, \qquad \deg\Phi(x)= 2,\vphantom{\Big[}
\eqno(4.1)
$$
and satisfy $\vphantom{\bigg[}$
$$
\eqalignno{&\hbox to 1.8truecm{$j(f)\phi(x)$\hfill}
\hbox{$=f(x)$\hfill},\vphantom{\Big[}&(4.2)\cr
&\hbox to 1.8truecm{$l(f)\phi(x)$\hfill}
\hbox{$=0$\hfill},\vphantom{\Big[}&\cr
&\hbox to 1.8truecm{$s\phi(x)$\hfill}
\hbox{$=\Phi(x)-{1\over 2}
\varpi^{ij}(x)\partial_i\phi(x)\partial_j\phi(x)
+( X^*{}^i+\varpi^{ij}(x)y_j)\partial_i\phi(x)$\hfill},
\vphantom{\Big[}&\cr
&\hbox to 1.8truecm{$j(f)\Phi(x)$\hfill}
\hbox{$=0$\hfill},\vphantom{\Big[}&\cr
&\hbox to 1.8truecm{$l(f)\Phi(x)$\hfill}
\hbox{$=0$\hfill},\vphantom{\Big[}&\cr
&\hbox to 1.8truecm{$s\Phi(x)$\hfill}
\hbox{$=-\varpi^{ij}(x)\partial_i\phi(x)\partial_j\Phi(x)
+( X^*{}^i+\varpi^{ij}(x)y_j)\partial_i\Phi(x)$\hfill},\vphantom{\Big[}&\cr}
$$
for any function $f$ in $\Fun(M)$.

From the discussion of sect. 1, we know that the image by $\varrho$ 
of the Hamilton--Weil generators $\phi(x)$, $\Phi(x)$ equals the image by 
$\varrho{}^\vee$ of the Poisson--Weil generators $\omega^i(x)$, 
$\Omega^i(x)$, respectively. Explicitly, denoting these objects by   
$\omega_\phi{}^i(x)$, $\Omega_\Phi{}^i(x)$, one has
$$
\eqalignno{\omega_\phi{}^i(x)=&-\varpi^{ij}(x)\partial_j\phi(x),
\vphantom{\Big[}&(4.3)\cr
\Omega_\Phi{}^i(x)=&-\varpi^{ij}(x)\partial_j\Phi(x).\vphantom{\Big[}&\cr}
$$
Using (4.1), (4.2), it is easy to check that $\omega_\phi{}^i(x)$, 
$\Omega_\Phi{}^i(x)$ fulfill (3.9), (3.10) and satisfy (3.11) 
with $j(u)$, $l(u)$, $u^i$, $\omega^i(x)$, $\Omega^i(x)$ substituted by 
$j(f)$, $l(f)$, $u_f{}^i$, $\omega_\phi{}^i(x)$, $\Omega_\Phi{}^i(x)$, 
respectively, for any function $f$ of $\Fun(M)$.

The equivariant extension of the Hamilton differential operation of $\Pi TM$,
(cfr. sect. 1), referred to as Hamilton  
equivariant differential operation below,  
is now easily obtained. It is the $\Fun(M)$ operation over 
$\Fun(\Pi T\Pi T^*M)\hat\otimes W(\Fun(M))$ with 
natural generators $x^i$, $\tilde X^i$, $y_i$, $\tilde Y_i$,
$\phi(x)$, $\Phi(x)$ satisfying (3.12) with $j(u)$, $l(u)$, 
$u^i$, $\omega^i(x)$, $\Omega^i(x)$ substituted by $j(f)$, $l(f)$, $u_f{}^i$, 
$\omega_\phi{}^i(x)$, $\Omega_\Phi{}^i(x)$, respectively, and 
$$
\eqalignno{&\hbox to 1.8truecm{$j(f)\phi(x)$\hfill}
\hbox{$=f(x)$\hfill},\vphantom{\Big[}&(4.4)\cr
&\hbox to 1.8truecm{$l(f)\phi(x)$\hfill}
\hbox{$=0$\hfill},\vphantom{\Big[}&\cr
&\hbox to 1.8truecm{$s\phi(x)$\hfill}
\hbox{$=\Phi(x)+{1\over 2}
\varpi^{ij}(x)\partial_i\phi(x)\partial_j\phi(x)
+(\tilde X^i+\varpi^{ij}(x)y_j)\partial_i\phi(x)$\hfill},
\vphantom{\Big[}&\cr
&\hbox to 1.8truecm{$j(f)\Phi(x)$\hfill}
\hbox{$=0$\hfill},\vphantom{\Big[}&\cr
&\hbox to 1.8truecm{$l(f)\Phi(x)$\hfill}
\hbox{$=0$\hfill},\vphantom{\Big[}&\cr
&\hbox to 1.8truecm{$s\Phi(x)$\hfill}
\hbox{$=(\tilde X^i+\varpi^{ij}(x)y_j)\partial_i\Phi(x)$\hfill},
\vphantom{\Big[}&\cr}
$$
for any function $f$ of $\Fun(M)$.
The Hamilton equivariant generators $\tilde X^i$, $\tilde Y_i$
are related the the Hamilton generators $X^*{}^i$, $Y^*{}_i$
by (3.14) with $\omega^i(x)$, $\Omega^i(x)$ 
substituted by $\omega_\phi{}^i(x)$, $\Omega_\Phi{}^i(x)$, respectively.
Obviously, $\omega_\phi{}^i(x)$, $\Omega_\Phi{}^i(x)$ satisfy (3.13) 
with $j(u)$, $l(u)$, $u^i$, $\omega^i(x)$, $\Omega^i(x)$ substituted by 
$j(f)$, $l(f)$, $u_f{}^i$, $\omega_\phi{}^i(x)$, $\Omega_\Phi{}^i(x)$, 
respectively, for any function $f$ of $\Fun(M)$, as usual.

As the above construction is completely local, it works also
for the local Hamilton symmetry at the price of having multivalued
Hamilton--Weil generators $\phi(x)$, $\Phi(x)$. This may be relevant 
in the analysis of the implications of the global topology of $M$.

\vskip .4cm
{\bf 5. The Differential $d$}
\vskip .4cm
\par
There is an important operator $d$ which enters the construction 
of the topological observables of the Poisson sigma model. For the sake of 
clarity, we shall analyze its properties separately in this section.

$d$ is the degree $+1$ derivation on $\Fun(\Pi T\Pi T^*M)\hat\otimes 
W(\Pois(M))$ defined in terms of the Poisson equivariant
generators by
$$
\eqalignno{&\hbox to 1.3truecm{$dx^i$\hfill}
\hbox{$=\tilde X^i$\hfill},\vphantom{\Big[}&(5.1)\cr
&\hbox to 1.3truecm{$d\tilde X^i$\hfill}
\hbox{$=0$\hfill},\vphantom{\Big[}&\cr
&\hbox to 1.3truecm{$dy_i$\hfill}
\hbox{$=\tilde Y_i$\hfill},\vphantom{\Big[}&\cr
&\hbox to 1.3truecm{$d\tilde Y_i$\hfill}
\hbox{$=0$\hfill},\vphantom{\Big[}&\cr
&\hbox to 1.3truecm{$d\omega^i(x) $\hfill}
\hbox{$=\Omega^i(x)+\tilde X^j\partial_j\omega^i(x)$\hfill}
\vphantom{\Big[},&(5.2)\cr
&\hbox to 1.3truecm{$d\Omega^i(x)$\hfill}
\hbox{$=\tilde X^j\partial_j\Omega^i(x)$\hfill}.\vphantom{\Big[}&\cr}
$$
The interest of $d$ stems from the fact that it is nilpotent and 
(anti)commutes with all the derivations of the Poisson 
equivariant operation of $\Pi TM$:
$$
\eqalignno{&\hbox to 1.5truecm{$[d,d]$\hfill}
\hbox{$=0$\hfill},\vphantom{\Big[}&(5.3)\cr
&\hbox to 1.5truecm{$[d,j(u)]$\hfill}
\hbox{$=0$\hfill},\vphantom{\Big[}&\cr
&\hbox to 1.5truecm{$[d,l(u)]$\hfill}
\hbox{$=0$\hfill},\vphantom{\Big[}&\cr
&\hbox to 1.5truecm{$[d,s]$\hfill}
\hbox{$=0$\hfill},\vphantom{\Big[}&\cr}
$$
for any Poisson vector $u^i$ field in $\Pois(M)$.

$d$ can be defined also in $\Fun(\Pi T\Pi T^*M)\hat\otimes W(\Fun(M))$
in terms of the Hamilton equivariant generators by the same relations (5.1)
and by
$$
\eqalignno{
&\hbox to 1.2truecm{$d\phi(x)$\hfill}
\hbox{$=\Phi(x)+\tilde X^i\partial_i\phi(x)$\hfill},\vphantom{\Big[}&(5.4)\cr
&\hbox to 1.2truecm{$d\Phi(x)$\hfill}
\hbox{$=\tilde X^i\partial_i\Phi(x)$\hfill}.\vphantom{\Big[}&\cr}
$$
(5.3) holds also in this case but with $j(u)$, $l(u)$ replaced by $j(f)$, 
$l(f)$, for any function $f$ in $\Fun(M)$.

It is easy to check, using (5.4), that $\omega_\phi{}^i(x)$, 
$\Omega_\Phi{}^i(x)$, given by (4.3), satisfy relations (5.2). 

\vskip .4cm
{\bf 6. Mod $d$ Poisson and Hamilton equivariant classes of $\Pi TM$}
\vskip .4cm
\par

To construct topological observables of the Poisson sigma model, 
one needs representatives ${\cal O}$ of mod $d$ Poisson 
equivariant classes. By definition, any such ${\cal O}$ is an element
of $\Fun(\Pi T\Pi T^*M)\hat\otimes W(\Pois(M))$ satisfying
$$
\eqalignno{
&\hbox to 1.2truecm{$j(u){\cal O}$\hfill}\hbox{$=d{\cal O}_{-1}(u)$,\hfill}
\vphantom{\Big[}&(6.1)\cr
&\hbox to 1.2truecm{$l(u){\cal O}$\hfill}\hbox{$=d{\cal O}_0(u)$,\hfill}
\vphantom{\Big[}&\cr
&\hbox to 1.2truecm{$s{\cal O}$\hfill}\hbox{$=d{\cal O}_{+1}$,\hfill}
\vphantom{\Big[}&\cr}
$$ 
for some ${\cal O}_{-1}(u)$, ${\cal O}_0(u)$, ${\cal O}_{+1}$ in 
$\Fun(\Pi T\Pi T^*M)\hat\otimes W(\Pois(M))$, for any Poisson vector 
$u^i$ field in $\Pois(M)$. An analogous definition holds
when restricting to the Hamilton symmetry
with ${\cal O}$, ${\cal O}_{-1}(u)$, ${\cal O}_0(u)$, ${\cal O}_{+1}$
substituted by elements ${\cal O}$, ${\cal O}_{-1}(f)$, ${\cal O}_0(f)$, 
${\cal O}_{+1}$ of $\Fun(\Pi T\Pi T^*M)\hat\otimes W(\Fun(M))$, for 
any function $f$ in $\Fun(M)$.

Let  $j_M$, $l_M$ and  $d_M$ denote the usual differential geometric 
contraction, Lie derivative and de Rham differential operators of $M$. 

Let $\beta^{i_1\cdots i_p}$ be any $p$--vector, which we represent 
in $\Fun(\Pi T\Pi T^*M)\hat\otimes W(\Pois(M))$ as usual as
$$
\beta(x,y)=\hbox{$1\over p!$}\beta^{i_1\cdots i_p}(x)y_{i_1}\cdots y_{i_p}.
\vphantom{\Big[}\eqno(6.2)
$$
Using (3.12), by a simple calculation, one finds
$$
\eqalignno{
&\hbox to 2.1truecm{$j(u)\beta(x,y)$\hfill}\hbox{$=0$,\hfill}
\vphantom{\Big[}&(6.3)\cr
&\hbox to 2.1truecm{$l(u)\beta(x,y)$\hfill}\hbox{$=l_M(u)\beta(x,y)$,\hfill}
\vphantom{\Big[}&\cr
&\hbox to 2.1truecm{$s\beta(x,y)$\hfill}
\hbox{$=d\beta(x,y)-[\varpi,\beta](x,y)+l_M(\omega)\beta(x,y)$,\hfill}
\vphantom{\Big[}&\cr}
$$ 
for any Poisson vector field $u^i$ in $\Pois(M)$.
If $\beta(x,y)$ is a representative of a Poisson invariant 
Poisson--Lichnerowicz cohomology class, i. e. 
$$
l_M(u)\beta(x,y)=0,\quad q\beta(x,y)=0,\vphantom{\Big[}
\eqno(6.4)
$$
for any Poisson vector field $u^i$ in $\Pois(M)$ (cfr. sect. 2, eq. (2.9)),
then
$$\eqalignno{
&\hbox to 2.2truecm{$j(u)\beta(x,y)$\hfill}\hbox{$=0$,\hfill}
\vphantom{\Big[}&(6.5)\cr
&\hbox to 2.2truecm{$l(u)\beta(x,y)$\hfill}\hbox{$=0$,\hfill}
\vphantom{\Big[}&\cr
&\hbox to 2.2truecm{$s\beta(x,y)$\hfill}\hbox{$=d\beta(x,y)$,\hfill}
\vphantom{\Big[}&\cr}
$$
for all with $u^i$ in $\Pois(M)$.
Thus, $\beta(x,y)$ is a representative of a mod $d$ Poisson equivariant 
cohomology class.

Let $\sigma_{i_1\cdots i_p}$ be any $p$--form, which we represent 
in $\Fun(\Pi T\Pi T^*M)\hat\otimes W(\Pois(M))$ as
$$
\sigma(x,\tilde X)=\hbox{$1\over p!$}\sigma_{i_1\cdots i_p}(x)
\tilde X^{i_1}\cdots \tilde X^{i_p}.\vphantom{\Big[}
\eqno(6.6)
$$
Using (3.12) again, one finds
$$
\eqalignno{&\hbox to 2.25truecm{$j(u)\sigma(x,\tilde X)$\hfill}
\hbox{$=0$\hfill},\vphantom{\Big[}&(6.7)\cr
&\hbox to 2.25truecm{$l(u)\sigma(x,\tilde X)$\hfill}
\hbox{$=l_M(u)\sigma(x,\tilde X)$\hfill},\vphantom{\Big[}&\cr
&\hbox to 2.25truecm{$s\sigma(x,\tilde X)$\hfill}
\hbox{$=d_M\sigma(x,\tilde X)+l_M(\omega)\sigma(x,\tilde X)
-j_M(\Omega)\sigma(x,\tilde X)$\hfill},\vphantom{\Big[}&\cr
&\hbox to 2.25truecm{$\hphantom{s\sigma(x,\tilde X)}$\hfill}
\hbox{$\hphantom{=}
+k d_M\sigma(x,\tilde X)-dk\sigma(x,\tilde X)$\hfill},\vphantom{\Big[}&\cr}
$$
for every Poisson vector field $u^i$ in $\Pois(M)$,
where the operator $k$ is the degree $0$ derivation defined by
$$
k\tilde X^i=\varpi^{ij}(x)y_j\vphantom{\Big[}
\eqno(6.8)
$$
and acting trivially on all other Poisson equivariant generators.
Therefore, if $\sigma(x,\tilde X)$ is a representative of a 
Poisson basic de Rham cohomology class, i. e.  
$$
j_M(u)\sigma(x,\tilde X)=0,\quad l_M(u)\sigma(x,\tilde X)=0,
\quad d_M\sigma(x,\tilde X)=0,\vphantom{\Big[}
\eqno(6.9)
$$
for any Poisson vector field $u^i$ in $\Pois(M)$, then
$$
\eqalignno{
&\hbox to 2.25truecm{$j(u)\sigma(x,\tilde  X) $\hfill}\hbox{$=0$,\hfill}
\vphantom{\Big[}&(6.10)\cr
&\hbox to 2.25truecm{$l(u)\sigma(x,\tilde  X)$\hfill}\hbox{$=0$,\hfill}
\vphantom{\Big[}&\cr
&\hbox to 2.25truecm{$s\sigma(x,\tilde  X)$\hfill}
\hbox{$=-dk\sigma(x,\tilde X)$,\hfill}
\vphantom{\Big[}&\cr}
$$ 
for all with $u^i$ in $\Pois(M)$.
Thus, $\sigma(x,\tilde  X)$ is a representative of a mod $d$ Poisson 
equivariant cohomology class.

Demanding invariance or basicity under the Poisson symmetry is very 
restrictive and in general only trivial or uninteresting solutions of
this requirement are available on a generic Poisson manifold. 
So, it is important to see whether restricting to the Hamilton symmetry 
yields mod $d$ Hamilton equivariant cohomology classes other than those 
obtained from the mod $d$ Poisson equivariant ones via pull--back
by the homomorphism $\varrho:\Fun(M)\rightarrow\Pois(M)$ (cfr. sect. 4). 

Consider again a $p$--vector $\beta^{i_1\cdots i_p}$ and view
$\beta(x,y)$ as an element of $\Fun(\Pi T\Pi T^*M)$
$\hat\otimes W(\Fun(M))$. Proceeding as in (6.3), one finds
$$
\eqalignno{
&\hbox to 2.1truecm{$j(f)\beta(x,y)$\hfill}\hbox{$=0$,\hfill}
\vphantom{\Big[}&(6.11)\cr
&\hbox to 2.1truecm{$l(f)\beta(x,y)$\hfill}
\hbox{$=[[\varpi,\beta],f](x,y)-[\varpi,[\beta,f]](x,y)$,\hfill}
\vphantom{\Big[}&\cr
&\hbox to 2.1truecm{$s\beta(x,y)$\hfill}
\hbox{$=d\beta(x,y)-[\varpi,\beta](x,y)
+[[\varpi,\beta],\phi](x,y)-[\varpi,[\beta,\phi]](x,y)$,\hfill}
\vphantom{\Big[}&\cr}
$$ 
for any function $f$ in $\Fun(M)$.
If $\beta(x,y)$ satisfies
$$
[f,\beta](x,y)=0,\quad q\beta(x,y)=0,\vphantom{\Big[}
\eqno(6.12)
$$
for any function $f$ in $\Fun(M)$, and is therefore a representative of a 
Hamilton invariant Poisson--Lichnerowicz cohomology class
(cfr. sect. 2, eq. (2.9)), then
$$\eqalignno{
&\hbox to 2.2truecm{$j(f)\beta(x,y)$\hfill}\hbox{$=0$,\hfill}
\vphantom{\Big[}&(6.13)\cr
&\hbox to 2.2truecm{$l(f)\beta(x,y)$\hfill}\hbox{$=0$,\hfill}
\vphantom{\Big[}&\cr
&\hbox to 2.2truecm{$s\beta(x,y)$\hfill}\hbox{$=d\beta(x,y)$,\hfill}
\vphantom{\Big[}&\cr}
$$
for all $f$ in $\Fun(M)$. Thus, $\beta(x,y)$ is a 
representative of a mod $d$ Hamilton equivariant cohomology class.

Consider again a $p$--form $\sigma_{i_1\cdots i_p}$ and view
$\sigma(x,\tilde X)$ as an element of $\Fun(\Pi T\Pi T^*M)$
$\hat\otimes W(\Fun(M))$. Proceeding as in (6.7) and performing some simple 
rearrangements, one finds 
$$
\eqalignno{
&\hbox to 2.25truecm{$j(f)\sigma(x,\tilde  X)$\hfill}\hbox{$=0$,\hfill}
\vphantom{\Big[}&(6.14)\cr
&\hbox to 2.25truecm{$l(f)\sigma(x,\tilde  X)$\hfill}
\hbox{$=dj_M(u_f)\sigma(x,\tilde X)+j_M(u_f)d_M\sigma(x,\tilde X)$,\hfill}
\vphantom{\Big[}&\cr
&\hbox to 2.25truecm{$s\sigma(x,\tilde  X)$\hfill}
\hbox{$=d\big(\sigma(x,\tilde X)-h\sigma(x,\tilde X)\big)
+h d_M\sigma(x,\tilde X)$,\hfill}
\vphantom{\Big[}&\cr}
$$ 
for every function $f$ in $\Fun(M)$, where the operator $h$ is the degree $0$ 
derivation defined by
$$
h\tilde  X^i=\varpi^{ij}(x)(y_j-\partial_j\phi(x))\vphantom{\Big[}
\eqno(6.15)
$$
and acting trivially on all other Hamilton equivariant generators
and $u_f{}^i$ is defined in (2.7).
Therefore, if $\sigma$ satisfies the condition
$$
k d_M\sigma(x,\tilde X)=0,\vphantom{\Big[}
\eqno(6.16)
$$
then
$$
\eqalignno{
&\hbox to 2.25truecm{$j(f)\sigma(x,\tilde  X)$\hfill}\hbox{$=0$,\hfill}
\vphantom{\Big[}&(6.17)\cr
&\hbox to 2.25truecm{$l(f)\sigma(x,\tilde  X)$\hfill}
\hbox{$=dj_M(u_f)\sigma(x,\tilde X)$,\hfill}
\vphantom{\Big[}&\cr
&\hbox to 2.25truecm{$s\sigma(x,\tilde  X)$\hfill}
\hbox{$=d\big(\sigma(x,\tilde X)-h\sigma(x,\tilde X)\big)$,\hfill}
\vphantom{\Big[}&\cr}
$$ 
for all $f$ in $\Fun(M)$.
Thus, $\sigma(x,\tilde  X)$ is a representative of a mod $d$ Hamilton
equivariant cohomology class.

\eject
\vskip .4cm
{\bf 7. Poisson and Hamilton Action of a Lie Algebra $\goth h$}
\vskip .4cm
\par

In the analysis of the Poisson sigma model expounded in later sections,
it turns out to be natural to restrict the symmetry Lie algebra 
to be some finite dimensional Lie subalgebra of the Poisson or Hamilton 
vector field Lie algebras.
This can be done efficiently by using the formalism of Poisson or Hamilton 
actions on $M$ of some abstract finite dimensional Lie algebra $\goth h$.

Let $\goth h$ be a Lie algebra and let $\{t_a\}$ be a basis of $\goth h$. 
Then, 
$$
[t_a,t_b]=c^c{}_{ab}t_c,\vphantom{\Big[}
\eqno(7.1)
$$
$c^c{}_{ab}$ being the structure constants of $\goth h$.

A Poisson (Hamilton) action of $\goth h$ on $M$ is a Lie algebra 
homomorphism $\upsilon:\goth h\rightarrow\Pois(M)$ 
($\varsigma:\goth h\rightarrow\Fun(M)$). 
In the Poisson case, $\upsilon(\goth h)$ is a Lie subalgebra of 
$\Pois(M)$. Indeed, setting $v_a{}^i=\upsilon^i(t_a)$, one has 
$$
[v_a,v_b]=c^c{}_{ab}v_c.\vphantom{\Big[}
\eqno(7.2)
$$
Similarly, in the Hamilton case, $\varsigma(\goth h)$ is a Lie subalgebra of 
$\Fun(M)$. Setting $h_a=\varsigma(t_a)$, one has 
$$
\{h_a,h_b\}=c^c{}_{ab}h_c.\vphantom{\Big[}
\eqno(7.3)
$$

The $\goth h$ Poisson (Hamilton) differential operation of $\Pi TM$ is the 
pull--back of the Poisson (Hamilton) differential operation of $\Pi TM$ by the 
Lie algebra homomorphism $\upsilon$ ($\varsigma$) (cfr. sects. 1, 3, 4). 
Hence, it is the $\goth h$ operation over $\Fun(\Pi T\Pi T^*M)$ 
with natural generators $x^i$, $X^*{}^i$, $y_i$, $Y^*{}_i$ 
satisfying (3.6) with $j(u)$, $l(u)$, $u^i$ ($j(f)$, $l(f)$, $f$) 
substituted by $j(r)$, $l(r)$, $\upsilon^i(r)$ ($u_{\varsigma(r)}{}^i$), 
respectively, for any element $r$ of $\goth h$. The $\goth h$ 
Poisson (Hamilton) generators $X^*{}^i$, $Y^*{}_i$ are 
related to the fundamental generators $X^i$, $Y_i$ again by $(3.7)$. 

Next, we consider the Weil operation of the Lie algebra $\goth h$ 
(cfr. sect. 1). It is the $\goth h$ operation over the Weil algebra
$W(\goth h)$ with natural generators $\gamma^a$, $\Gamma^a$ dual to the basis 
vector $t_a$ of degrees 
$$
\deg\gamma^a=1, \qquad \deg\Gamma^a=2,\vphantom{\Big[}
\eqno(7.4)
$$
and satisfying
$$
\eqalignno{&\hbox to 1.4truecm{$j(r)\gamma^a$\hfill}
\hbox{$=r^a$\hfill},\vphantom{\Big[}&(7.5)\cr
&\hbox to 1.4truecm{$l(r)\gamma^a$\hfill}
\hbox{$=-c^a{}_{bc}r^b\gamma^c$\hfill},\vphantom{\Big[}&\cr
&\hbox to 1.4truecm{$s\gamma^a$\hfill}
\hbox{$=\Gamma^a-{1\over 2}c^a{}_{bc}\gamma^b\gamma^c$\hfill},
\vphantom{\Big[}&\cr
}
$$
$$
\eqalignno{
&\hbox to 1.4truecm{$j(r)\Gamma^a$\hfill}
\hbox{$=0$\hfill},\vphantom{\Big[}&\cr
&\hbox to 1.4truecm{$l(r)\Gamma^a$\hfill}
\hbox{$=-c^a{}_{bc}r^b\Gamma^c$\hfill},\vphantom{\Big[}&\cr
&\hbox to 1.4truecm{$s\Gamma^a$\hfill}
\hbox{$=-c^a{}_{bc}\gamma^b\Gamma^c$\hfill},\vphantom{\Big[}&\cr}
$$
with $r$ in $\goth h$.

From the discussion of sect. 1, we know that the image by $\upsilon$
($\varsigma$) of the $\goth h$ Weil generators 
$\gamma^a$, $\Gamma^a$ equals the image 
by $\upsilon^\vee$ ($\varsigma^\vee$) of the Poisson--Weil (Hamilton--Weil) 
generators $\omega^i(x)$, $\Omega^i(x)$, ($\phi(x)$, $\Phi(x)$), 
respectively. Explicitly, denoting these objects by   
$\omega_\gamma{}^i(x)$, $\Omega_\Gamma{}^i(x)$, 
($\phi_\gamma(x)$, $\Phi_\Gamma(x)$), one has
$$
\eqalignno{&\hbox to 1.35truecm{$\omega_\gamma{}^i(x)$\hfill}
\hbox to 3.5truecm{$=\sum_a\gamma^av_a{}^i(x)$,\hfill}
\hbox to 1.35truecm{$\Omega_\Gamma{}^i(x)$\hfill}
\hbox to 3.5truecm{$=\sum_a\Gamma^av_a{}^i(x)$.\hfill}\vphantom{\Big[}&(7.6)\cr
&\hbox to 1.35truecm{$(\phi_\gamma(x)$\hfill}
\hbox to 3.5truecm{$=\sum_a\gamma^ah_a(x)$,\hfill}
\hbox to 1.35truecm{$\Phi_\Gamma(x)$\hfill}
\hbox to 3.5truecm{$=\sum_a\Gamma^ah_a(x).)$\hfill}\vphantom{\Big[}&(7.7)\cr}
$$

Using (7.4), (7.5), it is easy to check that $\omega_\gamma{}^i(x)$, 
$\Omega_\Gamma{}^i(x)$ ($\phi_\gamma(x)$, $\Phi_\Gamma(x)$)
fulfill (3.9), (3.10) ((4.1)) and satisfy (3.11) ((4.2)) with 
$j(u)$, $l(u)$, $u^i$,  $\omega^i(x)$, $\Omega^i(x)$ 
($j(f)$, $l(f)$, $f$, $\phi(x)$, $\Phi(x)$)
substituted by $j(r)$, $l(r)$, $\upsilon^i(r)$, $\omega_\gamma{}^i(x)$,
$\Omega_\Gamma{}^i(x)$ ($u_{\varsigma(r)}{}^i$, $\phi_\gamma(x)$, 
$\Phi_\Gamma(x)$), respectively, for $r$ in $\goth h$.

The equivariant extension of the $\goth h$ Poisson (Hamilton) 
differential operation of $\Pi TM$ (cfr. sect. 1), which we shall call 
$\goth h$ Poisson (Hamilton) equivariant differential operation below,
is now easily obtained. It is the $\goth h$ operation over 
$\Fun(\Pi T\Pi T^*M)\hat\otimes W(\goth h)$ with 
natural generators $x^i$, $\tilde X^i$, $y_i$, $\tilde Y_i$,
$\gamma^a$, $\Gamma^a$ satisfying (3.12), with $j(u)$, $l(u)$, 
$u^i$, $\omega^i(x)$, $\Omega^i(x)$ ($j(f)$, $l(f)$, $f$, 
$\omega_\phi{}^i(x)$, $\Omega_\Phi{}^i(x)$)
substituted by $j(r)$, $l(r)$, $\upsilon^i(r)$, $\omega_\gamma{}^i(x)$,
$\Omega_\Gamma{}^i(x)$ ($u_{\varsigma(r)}{}^i$, 
$\omega_{\phi_\gamma}{}^i(x)$, $\Omega_{\Phi_\Gamma}{}^i(x)$), 
respectively, and (7.5), for any element $r$ of $\goth h$.
The $\goth h$ Poisson (Hamilton) equivariant generators $
\tilde X^i$, $\tilde Y_i$ are related the the $\goth h$ Poisson (Hamilton) 
generators $X^*{}^i$, $Y^*{}_i$ by (3.14) with $\omega^i(x)$, $\Omega^i(x)$ 
($\omega_\phi{}^i(x)$, $\Omega_\Phi{}^i(x)$) substituted by 
$\omega_\gamma{}^i(x)$, $\Omega_\Gamma{}^i(x)$ 
($\omega_{\phi_\gamma}{}^i(x)$, $\Omega_{\Phi_\Gamma}{}^i(x)$)
(cfr. eq. (4.3)), respectively.

The $d$ operator is defined in obvious fashion
$$
\eqalignno{&\hbox to 0.8truecm{$d\gamma^a$\hfill}
\hbox{$=\Gamma^a$\hfill},\vphantom{\Big[}&(7.8)\cr
&\hbox to 0.8truecm{$d\Gamma^a$\hfill}
\hbox{$=0$\hfill}.\vphantom{\Big[}&\cr}
$$
(5.3) holds with $j(u)$, $l(u)$ substituted by $j(r)$, $l(r)$,
respectively, for $r$ in $\goth h$.

Representatives of mod $d$ $\goth h$ Poisson (Hamilton) equivariant classes 
are obtained from those of mod $d$ Poisson and Hamilton equivariant classes 
discussed in sect. 6 by pull-back via the Poisson (Hamilton) action $\upsilon$ 
($\varsigma$) of $\goth h$ on $M$.

\vskip .4cm
{\bf 8. 2--Dimensional de Rham Superfields and Singular Superchains}
\vskip .4cm
\par

In general, the fields of a 2--dimensional field theory are differential 
forms on a $2$--dimensional manifold $\Sigma$. They can be viewed as 
elements of the space $\Fun(\Pi T\Sigma)$ of functions on the parity 
reversed tangent bundle $\Pi T\Sigma$ of $\Sigma$, which we shall call 
de Rham superfields \ref{6}. 
More explicitly, we associate to the coordinates $z^\alpha$ of 
$\Sigma$ Grassmann odd partners $\zeta^\alpha$ with
$$
\deg z^\alpha=0, \qquad \deg\zeta^\alpha=1.\vphantom{\Big[}
\eqno(8.1)
$$
and
$$
dz^\alpha=\zeta^\alpha,\qquad d\zeta^\alpha=0.\vphantom{\Big[}
\eqno(8,2)
$$
A generic de Rham superfield $\Psi(z,\zeta)$ is a triplet 
formed by a $0$--, $1$-- and $2$--form field $\psi^{(0)}(z)$, 
$\psi^{(1)}{}_\alpha(z)$ and $\psi^{(2)}{}_{\alpha\beta}(z)$ organized as
$$
\Psi(z,\zeta)=\psi^{(0)}(z)+\zeta^\alpha\psi^{(1)}{}_\alpha(z)
+\hbox{$1\over 2$}\zeta^\alpha\zeta^\beta\psi^{(2)}{}_{\alpha\beta}(z).
\eqno(8.3)
$$
Note that in this formalism, the de Rham differential $d$ of $\Sigma$
is simply
$$
d=\zeta^\alpha\partial/\partial z^\alpha.
\eqno(8.4)
$$

The coordinate invariant integration measure is 
$$
\mu={\rm d}z^1{\rm d}z^2{\rm d}\zeta^1{\rm d}\zeta^2.
\eqno(8.5)
$$
Any de Rham superfield $\Psi$ can be integrated on $\Sigma$ according to
the prescription
$$
\int_\Sigma\mu\Psi=\int_\Sigma\hbox{$1\over 2$}
dz^\alpha dz^\beta\psi^{(2)}{}_{\alpha\beta}(z).
\eqno(8.6)
$$
By Stokes' theorem 
$$
\int_\Sigma\mu d\Psi=0.
\eqno(8.7)
$$

The singular chain complex of $\Sigma$ can be given a parallel treatment. 
A singular superchain $C$ is a triplet formed by a $0$--, 
$1$-- and $2$--dimensional singular chain $C_{(0)}$, $C_{(1)}$, $C_{(2)}$
organized as a formal sum
$$
C=C_{(0)}+C_{(1)}+C_{(2)}.
\eqno(8.8)
$$
The singular boundary operator $\partial$ extends to superchains in obvious 
fashion by setting
$$
(\partial C)_{(0)}=\partial C_{(1)}, 
\quad (\partial C)_{(1)}=\partial C_{(2)},
\quad (\partial C)_{(2)}=0.
\eqno(8.9)
$$
A singular supercycle $Z$ is a superchain such that 
$$
\partial Z=0.
\eqno(8.10)
$$

A de Rham superfield $\Psi$ can be integrated on a superchain $C$:
$$
\int_C\mu\Psi=\int_{C_{(0)}}\psi^{(0)}
+\int_{C_{(1)}}dz^\alpha\psi^{(1)}{}_\alpha(z)
+\int_{C_{(2)}}\hbox{$1\over 2$}
dz^\alpha dz^\beta\psi^{(2)}{}_{\alpha\beta}(z).
\eqno(8.11)
$$
Stokes' theorem states that 
$$
\int_C\mu d\Psi=\int_{\partial C}\mu \Psi.
\eqno(8.12)
$$
In particular, if $Z$ is a supercycle,
$$
\int_Z\mu d\Psi=0.
\eqno(8.13)
$$

In the case where $\Sigma$ has a non empty boundary
$\partial\Sigma$, the above relations hold provided the component fields 
of the superfield obey suitable boundary conditions \ref{6}.

\vskip .4cm
{\bf 9. The $\goth h$ Poisson Sigma Model}
\vskip .4cm
\par

The Poisson Sigma Model is a 2--dimensional field theory whose base space is 
a closed $2$--dimensional surface $\Sigma$ and whose target space is a Poisson 
manifold $M$.

We assume that a finite dimensional  Lie algebra $\goth h$ is given 
together with a Hamilton action of $\goth h$ on $M$ 
$\varsigma:\goth h\rightarrow\Fun(M)$ (see sect. 7).
 
The fields of the $\goth h$ Poisson sigma model are organized in an operation, 
referred to as the $\goth h$ Hamilton de Rham superfield operation below. 
This is a de Rham superfield realization of the $\goth h$ Hamilton 
equivariant operation of $\Pi TM$ and is concretely constructed as follows.
Each of the generators $x^i$, $\tilde X^i$, $y_i$, $\tilde Y_i$, 
$\gamma^a$, $\Gamma^a$ of $\Fun(\Pi T\Pi T^*M)\hat\otimes W(\goth h)$ 
is realized as a de Rham superfield, denoted by the same symbol.
The values $x^i(z,\zeta)$, $\tilde X^i(z,\zeta)$, 
$y_i(z,\zeta)$, $\tilde Y_i(z,\zeta)$, $\gamma^a(z,\zeta)$, 
$\Gamma^a(z,\zeta)$ of these superfields for varying $(z,\zeta)$
generate, after imposing a natural smoothness requirement, 
a formal graded associative algebra ${\sh F}(\Sigma,M,\goth h)$. 
The $\goth h$ Hamilton de Rham superfield 
operation is the $\goth h$ operation over ${\sh F}(\Sigma,M,\goth h)$
whose derivations $j(r)$, $l(r)$, $r\in\goth h$, and $s$ 
are defined in terms of the de Rham superfield generators $x^i$, 
$\tilde X^i$, $y_i$, $\tilde Y_i$, $\gamma^a$, $\Gamma^a$ of 
${\sh F}(\Sigma,M,\goth h)$ according to expressions formally identical to 
those valid for the corresponding generators of $\Fun(\Pi T\Pi T^*M)
\hat\otimes W(\goth h)$, as expounded in sect. 7 above.
The derivation $d$ defined in (5.1), (7.8) is realized as the de Rham
differential $d$, eq. (8.4), as indicated by the use of the same notation.

A de Rham superfield $\Lambda$ is Hamilton, if $\Lambda(z,\zeta)$
belongs to ${\sh F}(\Sigma,M,\goth h)$ for all $(z,\zeta)$.
For any Hamilton de Rham superfield $\Lambda$, $\int_\Sigma\mu\Lambda$ is 
defined and belongs to ${\sh F}(\Sigma,M,\goth h)$. 

A Hamilton de Rham superfield $\Lambda$ is local if 
$\Lambda(z,\zeta)$ depends only on the values of the de Rham superfield 
generators $x^i$, $\tilde X^i$, $y_i$, $\tilde Y_i$, $\gamma^a$, 
$\Gamma^a$ and a finite number of their derivatives at $(z,\zeta)$. 
Clearly, each superfield $x^i$, $\tilde X^i$, $y_i$, $\tilde Y_i$, 
$\gamma^a$, $\Gamma^a$ is Hamilton and local.

If the Hamilton de Rham superfield $\Lambda$ is a representative of a 
mod $d$ $\goth h$ Hamilton de Rham superfield basic cohomology class, then 
$\int_\Sigma\mu \Lambda$ is a representative of a $\goth h$ Hamilton 
de Rham superfield basic cohomology class. Indeed, as $j(r)\Lambda$, 
$l(r)\Lambda$, $r\in\goth h$, and $s\Lambda$ all vanish mod $d$,
$j(r)\int_\Sigma\mu\Lambda$, $l(r)\int_\Sigma\mu \Lambda$, $r\in\goth h$, 
and $s\int_\Sigma\mu \Lambda$, vanish exactly on account of (8.7).

The crucial observations, which we shall exploit extensively below, 
are the following.
{\it Every element $\cal O$ of $\Fun(\Pi T\Pi T^*M)\hat\otimes W(\goth h)$
yields a local Hamilton de Rham superfield of ${\sh F}(\Sigma,M,\goth h)$, 
denoted also by $\cal O$ and called its Hamilton de Rham superfield 
realization, by substituting each of the 
generators of $\Fun(\Pi T\Pi T^*M)\hat\otimes W(\goth h)$ with the 
corresponding superfield generator of ${\sh F}(\Sigma,M,\goth h)$.  
Every relation involving one or more elements in $\Fun(\Pi T\Pi T^*M)
\hat\otimes W(\goth h)$ entails a formally identical relation involving their 
Hamilton de Rham superfield realizations in ${\sh F}(\Sigma,M,\goth h)$.
In particular, representatives of mod
$d$ $\goth h$ Hamilton equivariant cohomology classes yield directly
local Hamilton de Rham superfields representing mod $d$ $\goth h$ 
Hamilton de Rham superfield basic cohomology classes.}

The Lagrangian of the $\goth h$ Poisson sigma model is derived directly from 
the following degree $2$ element of $\Fun(\Pi T\Pi T^*M)
\hat\otimes W(\goth h)$ 
$$
{\cal L}_\pi=y_i\tilde X^i+\hbox{$1\over 2$}\pi^{ij}(x)y_iy_j-\Phi_\Gamma(x).
\eqno(9.1)
$$
Here, $\pi^{ij}$ is a $2$--vector satisfying 
$$
\eqalignno{&[\pi,\pi]=0,\,~~\hbox{i. e.}~~
\pi^{il}\partial_l\pi^{jk}+\pi^{jl}\partial_l\pi^{ki}
+\pi^{kl}\partial_l\pi^{ij}=0,\vphantom{\Big[}&(9.2)\cr
&[\pi,\varpi]=0,~~\hbox{i. e.}~~
\varpi^{il}\partial_l\pi^{jk}+\varpi^{jl}\partial_l\pi^{ki}
+\varpi^{kl}\partial_l\pi^{ij}\vphantom{\Big[}&(9.3)\cr
&\hphantom{[\pi,\varpi]=0,~~\hbox{i. e.}~~}
+\pi^{il}\partial_l\varpi^{jk}+\pi^{jl}\partial_l\varpi^{ki}
+\pi^{kl}\partial_l\varpi^{ij}=0.\vphantom{\Big[}&\cr}
$$
We further demand that the Hamilton action $\varsigma$ satisfies
$$
[\pi,\varsigma(r)]=0,~~\hbox{i. e.}~~\pi^{ij}\partial_j\varsigma(r)=0,
\eqno(9.4)
$$
with $r\in\goth h$. These restrictions on $\pi^{ij}$ and $\varsigma$, 
whose justification will be provided in the next section, have the following 
simple geometrical interpretation. $\pi^{ij}$ is another Poisson 
$2$--vector of $M$ compatible with the given Poisson $2$--vector 
$\varpi^{ij}$ (cfr. eq. (2.1)). The $\varsigma(r)$ are Casimir functions 
of the Poisson structure of $\pi^{ij}$. To avoid possible confusion,
{\it below, unless otherwise stated, we tacitly assume that the Poisson 
structure of $M$ is that defined by the Poisson $2$--vector $\varpi^{ij}$}. 

Let $\Cas_\pi(M)$ be the space of functions $f$ satisfying 
$$
[\pi,f]=0,~~\hbox{i. e.}~~ \pi^{ij}\partial_jf=0.
\eqno(9.5)
$$
Using (9.3) and the simple relation $\{f,g\}=[f,[\varpi,g]]$,
$f,~g\in\Fun(M)$, it is easy to show that $\Cas_\pi(M)$
a Poisson subalgebra of $\Fun(M)$, the ``$\pi$--twisted'' 
Casimir subalgebra. 

Using the relation $l_M(u_f)\pi=-[[\varpi,f],\pi]$, $f\in\Fun(M)$, 
it is simple to check that, for $f$ in $\Cas_\pi(M)$, $l_M(u_f)\pi^{ij}=0$. 
Hence, the Poisson $2$--vector $\pi^{ij}$ is invariant under the Hamilton 
vector fields of the $\pi$--twisted Casimir functions. 

From (9.4) and these simple considerations, it follows that, for $r\in\goth h$,
$\varsigma(r)\in\Cas_\pi(M)$ and that the Poisson $2$--vector 
$\pi^{ij}$ is invariant under the Hamilton action $\varsigma$, that is   
$l_M(u_{\varsigma(r)})\pi^{ij}$ $=0$, for $r\in\goth h$.

From here, using (9.3) and proceeding as in sect. 6, 
we find that ${\cal L}_\pi$ satisfies
$$
\eqalignno{
&\hbox to 1.3truecm{$j(r){\cal L}_\pi$\hfill}\hbox{$=0$,\hfill}
\vphantom{\Big[}&(9.6)\cr
&\hbox to 1.3truecm{$l(r){\cal L}_\pi$\hfill}\hbox{$=0$,\hfill}
\vphantom{\Big[}&\cr
&\hbox to 1.3truecm{$s{\cal L}_\pi$\hfill}
\hbox{$=d\big(y_i\tilde X^i-\Phi_\Gamma(x)
+\hbox{$1\over 2$}\pi^{ij}(x)y_iy_j
-\hbox{$1\over 2$}\varpi^{ij}(x)y_iy_j\big)$,\hfill}
\vphantom{\Big[}&\cr}
$$ 
for any element $r$ of $\goth h$. Hence, {\it ${\cal L}_\pi$ is a 
representative of a mod $d$ degree $2$ $\goth h$ Hamilton equivariant 
cohomology class}.
 
The treatment of the $\goth h$ Poisson sigma model requires going onto the 
$\goth h$ Hamilton de Rham superfield operation. The Lagrangian of the model 
is the local Hamilton de Rham superfield realization of ${\cal L}_\pi$
and is obtained from (9.1) using (5.1), (7.8). 
The action ${\cal S}_\pi$ of the model, given as usual
by $\int_\Sigma\mu {\cal L}_\pi$, thus reads explicitly
$$
{\cal S}_\pi=\int_\Sigma\mu\big(y_idx^i+\hbox{$1\over 2$}\pi^{ij}(x)y_iy_j
-\Phi_{d\gamma}(x)\big).
\eqno(9.7)
$$
So, ${\cal S}_\pi$ has degree $0$ and,
by (9.6) and the above discussion, satisfies
$$
\eqalignno{
&\hbox to 1.3truecm{$j(r){\cal S}_\pi$\hfill}\hbox{$=0$,\hfill}
\vphantom{\Big[}&(9.8)\cr
&\hbox to 1.3truecm{$l(r){\cal S}_\pi$\hfill}\hbox{$=0$,\hfill}
\vphantom{\Big[}&\cr
&\hbox to 1.3truecm{$s{\cal S}_\pi$\hfill}
\hbox{$=0$,\hfill}
\vphantom{\Big[}&\cr}
$$
with $r$ in $\goth h$. Thus, {\it ${\cal S}_\pi$ is a representative of 
a degree $0$ $\goth h$ Hamilton de Rham superfield basic cohomology class}.

If $\pi^{ij}=\varpi^{ij}$, $\Cas_\pi(M)=\Cas(M)$ on account of (2.5), (9.5).
Hence, the Poisson subalgebra $\varsigma(\goth h)$ is contained in the 
Casimir subalgebra $\Cas(M)$ and, as the Hamilton vector field of a Casimir 
function vanishes identically by (2.5), (2.7), 
the action of the derivations $j(r)$, $l(r)$ is trivial 
for all $r$ in $\goth h$. In this way, the underlying $\goth h$ Hamilton
equivariant cohomology of $\Pi TM$ reduces to ordinary cohomology
and, in this sense, is trivialized. 
When, conversely, $\pi^{ij}\not=\varpi^{ij}$, $\Cas_\pi(M)\not=\Cas(M)$ in 
general. Therefore, the above argument does not apply and 
the action of the derivations $j(r)$, $l(r)$ for $r$ in 
$\goth h$ is generally non trivial. In this way, the $\goth h$ Hamilton 
equivariant cohomology of $\Pi TM$ is generally non trivial as well. 
The import of this observation has been
discussed at the end of the introduction, sect. 0.

\vskip .4cm
{\bf 10. The Batalin--Vilkoviski Formulation of the $\goth h$
Poisson Sigma Model}
\vskip .4cm
\par
The superfield formulation of the Poisson sigma model was developed 
in order to implement the Batalin--Vilkoviski quantization algorithm
\ref{10}. It is encouraging to find out that the action ${\cal S}_\pi$
constructed above satisfies the Batalin--Vilkoviski 
classical master equation. 

Here, we shall use the convenient de Rham superfield formalism.
We identify the fields and antifields with $x^i$ and $y_i$, respectively.  
The Batalin--Vilkoviski odd symplectic form of the Poisson sigma model is 
$$
\widehat{\Omega}_{BV}=\int_\Sigma\mu\,\delta x^i\delta y_i.
\eqno(10.1)
$$
Note that there is no term corresponding to $\gamma^a$ and its antifield
in the symplectic form, since these
are considered fixed non dynamical background fields.

Therefore, the Batalin--Vilkoviski antibrackets are given by
$$
\big(x^i(z,\zeta),y_i(z',\zeta')\big)=\delta^i{}_j\delta(z,\zeta;z',\zeta'),
\eqno(10.2)
$$
where the super delta distribution $\delta$ is given by
$$
\delta(z,\zeta;z',\zeta')=
\hbox{$1\over 2$}\delta^{0,2}_{\alpha'\beta'}(z;z')
\zeta^{\alpha'}\zeta^{\beta'} 
+\delta^{1,1}_{\alpha\beta'}(z;z')\zeta^\alpha\zeta^{\beta'}
+\hbox{$1\over 2$}\delta^{2,0}_{\alpha\beta}(z;z')
\zeta^\alpha\zeta^\beta,
\eqno(10.3)
$$
$\delta^{p,1-p}(z;z')$ being the usual delta distributions for forms
on $\Sigma$. For a superfield $\Psi$
$$
\int_\Sigma\mu'\,\delta(z,\zeta;z',\zeta')\Psi(z',\zeta')=
\Psi(z,\zeta).
\eqno(10.4)
$$

Using (10.2), (10.4), one verifies that 
$$
\big({\cal S}_\pi,{\cal S}_\pi\big)=
\int_\Sigma\mu\big[2\pi{}^{ij}(x)y_i\partial_j\Phi_{d\gamma}(x)
-\hbox{$1\over 3$}\big(\pi^{il}\partial_l\pi^{jk}+\pi^{jl}\partial_l\pi^{ki}
+\pi^{kl}\partial_l\pi^{ij}\big)(x)y_iy_jy_k\big].
\eqno(10.5)
$$
Hence, {\it the action ${\cal S}_\pi$ satisfies the Batalin--Vilkoviski
classical master equation}
$$ 
\big({\cal S}_\pi,{\cal S}_\pi\big)=0,
\eqno(10.6)
$$
{\it if (9.2) and (9.4) hold}. This analysis provides a field theoretic 
justification of conditions (9.2) and (9.4). 

The field equations entailed by the action ${\cal S}_\pi$ are 
$$
\eqalignno{
&dx^i+\pi{}^{ij}(x)y_j=0,\vphantom{\Big[}\vphantom{\Big[}&(10.7)\cr
&dy_i+\hbox{$1\over 2$}\partial_i\pi{}^{jk}(x)y_jy_k
-\partial_i\Phi_{d\gamma}(x)=0.\vphantom{\Big[}&\cr}
$$
By applying the differential $d$ to both equations, one obtains the 
integrability condition  
$$
\pi{}^{ij}(x)\partial_j\Phi_{d\gamma}(x)
-\hbox{$1\over 2$}\big(\pi^{il}\partial_l\pi^{jk}+\pi^{jl}\partial_l\pi^{ki}
+\pi^{kl}\partial_l\pi^{ij}\big)(x)y_jy_k=0.\vphantom{\Big[}
\eqno(10.8)
$$
Hence, {\it the field equations are solvable if (9.2) and (9.4) hold.} 
It is interesting to note that the requirement of integrability of 
the field equations leads to the same restrictions as those implied by the 
master equations.

The Batalin--Vilkoviski variation of the superfields $x^i$, $y_i$ are given 
by
$$
\eqalignno{&\hbox to 0.9truecm{$\delta_\pi x^i$\hfill}
\hbox to 1.95truecm{$=(S_\pi,x^i)$\hfill}
\hbox{$=dx^i+\pi{}^{ij}(x)y_j$\hfill},\vphantom{\Big[}&(10.9)\cr
&\hbox to 0.9truecm{$\delta_\pi y_i$\hfill}
\hbox to 1.95truecm{$=(S_\pi,y_i)$\hfill}
\hbox{$=dy_i+{1\over 2}\partial_i\pi{}^{jk}(x)y_jy_k
-\partial_i\Phi_{d\gamma}(x)$\hfill}.\vphantom{\Big[}&\cr}
$$
As is well-known \ref{10}, if the master equation (10.6) is fulfilled, 
$\delta_\pi$ is a degree $1$ nilpotent derivation on 
${\sh F}(\Sigma,M,\goth h)$ 
$$
\delta_\pi{}^2=0.\vphantom{\Big[}
\eqno(10.10)
$$
From (5.1), (7.8), it is easy to see that $\delta_\pi$ is nothing but the 
Hamilton de Rham superfield realization of a degree $1$ derivation $w_\pi$ 
on $\Fun(\Pi T\Pi T^* M)\hat\otimes W(\goth h)$ defined by
$$
\eqalignno{
&\hbox to 1.12truecm{$w_\pi x^i$\hfill}
\hbox{$=\tilde X^i+\pi{}^{ij}(x)y_j$\hfill},\vphantom{\Big[}&(10.11)\cr
&\hbox to 1.12truecm{$w_\pi \tilde X^i$\hfill}
\hbox{$=-\pi{}^{ij}(x)\tilde Y_j
-\partial_j\pi{}^{ik}(x)\tilde X^jy_k$\hfill},\vphantom{\Big[}&\cr
}
$$
$$
\eqalignno{
&\hbox to 1.12truecm{$w_\pi y_i$\hfill}
\hbox{$=\tilde Y_i+{1\over 2}\partial_i\pi{}^{jk}(x)y_jy_k
-\partial_i\Phi_\Gamma(x)$\hfill},\vphantom{\Big[}&\cr
&\hbox to 1.12truecm{$w_\pi \tilde Y_i$\hfill}
\hbox{$=-{1\over 2}\partial_i\partial_j\pi{}^{kl}(x)
\tilde X^jy_ky_l+\partial_i\pi{}^{jk}(x)y_j\tilde Y_k
+ \partial_i\partial_j\Phi_\Gamma(x)\tilde X^j$\hfill},\vphantom{\Big[}&\cr
&\hbox to 1.12truecm{$w_\pi \gamma^a$\hfill}\hbox{$=0$\hfill},
\vphantom{\Big[}&\cr
&\hbox to 1.12truecm{$w_\pi \Gamma^a$\hfill}\hbox{$=0$\hfill}.
\vphantom{\Big[}&\cr}
$$
It is straightforward to verify that 
$$
\eqalignno{&\hbox to 1.77truecm{$[w_\pi,w_\pi]$\hfill}
\hbox{$=0$\hfill},\vphantom{\Big[}&(10.12)\cr
&\hbox to 1.77truecm{$[w_\pi,j(r)]$\hfill}
\hbox{$=0$\hfill},\vphantom{\Big[}&\cr
&\hbox to 1.77truecm{$[w_\pi,l(r)]$\hfill}
\hbox{$=0$\hfill},\vphantom{\Big[}&\cr
&\hbox to 1.77truecm{$[w_\pi,s]$\hfill}
\hbox{$=0$\hfill},\vphantom{\Big[}&\cr
&\hbox to 1.77truecm{$[w_\pi,d]$\hfill}
\hbox{$=0$\hfill},\vphantom{\Big[}&\cr}
$$
for $r\in\goth h$, {\it if (9.2), (9.3) and (9.4) hold.}
Hence, the compatibility of the nilpotent operator $w_\pi$ and 
the derivations of the $\goth h$ Hamilton equivariant operation of $\Pi TM$
leads to condition (9.3) in addition to conditions (9.2) and (9.4) 
previously obtained. 

\vskip .4cm
{\bf 11. $\goth h$ Hamilton de Rham Superfield Basic Cohomology Classes
and Batalin --Vilkoviski Observables}
\vskip .4cm
\par
Next, we want to investigate under which conditions local representatives 
of $\goth h$ Hamilton de Rham superfield basic cohomology classes are 
also Batalin--Vilkoviski observables of the $\goth h$ Poisson sigma model, 
i. e. local representatives of the $\delta_\pi$ cohomology classes \ref{10}. 

Let ${\cal O}$ be a local Hamilton de Rham superfield in 
${\sh F}(\Sigma,M,\goth h)$ representing a mod $d$ $\goth h$ Hamilton de Rham 
superfield basic cohomology class. Then, for any singular supercycle $Z$ 
(cfr. sect. 9), 
$$
\langle Z,{\cal O}\rangle=\int_Z\mu{\cal O}.
\eqno(11.1)
$$  
is a representative of a $\goth h$ Hamilton de Rham superfield 
basic cohomology class. Indeed, as $j(r){\cal O}$, 
$l(r){\cal O}$, $r\in\goth h$, and $s{\cal O}$ all vanish mod $d$,
$j(r)\int_Z\mu{\cal O}$, $l(r)\int_Z\mu {\cal O}$, $r\in\goth h$, 
and $s\int_Z\mu {\cal O}$, vanish exactly on account of (8.13).

According to the Batalin--Vilkoviski theory, $\langle Z,{\cal O}\rangle$ 
is an observable of the $\goth h$ Poisson sigma model
for all singular supercycles $Z$, provided
\eject
$$
\delta_\pi \langle Z,{\cal O}\rangle=0 
\eqno(11.2)
$$  
for all such $Z$ \ref{10}. This poses further restriction on ${\cal O}$,
namely
$$
\delta_\pi {\cal O}=d{\cal X},
\eqno(11.3)
$$ 
for some local Hamilton de Rham superfield ${\cal X}$ in 
${\sh F}(\Sigma,M,\goth h)$. 

Assume that ${\cal O}$ is the Hamilton de Rham superfield realization of 
some element of $\Fun(\Pi T\Pi T^* M)\hat\otimes W(\goth h)$, 
which we also denote by ${\cal O}$.  Then, on one hand $\cal O$ must obey 
(6.1) with $j(u)$, $l(u)$, $u^i$ replaced by $j(r)$, $l(r)$, $r$ with  
$r$ in $\goth h$. On the other, recalling that $\delta_\pi$ is 
the Hamilton de Rham superfield realization of $w_\pi$ (cfr. sect. 10), 
$\cal O$ must satisfy the further condition 
$$
w_\pi {\cal O}=d{\cal X},
\eqno(11.4)
$$ 
for some element ${\cal X}$ of $\Fun(\Pi T\Pi T^* M)\hat\otimes W(\goth h)$,
in analogy to (11.3).

Suppose ${\cal O}=\beta(x,y)$ is of the form (6.2). Then, $\beta(x,y)$ obeys
(6.12) with $f$ in $\varsigma(\goth h)$. Using (10.11), one computes
$$
w_\pi\beta(x,y)=
d\beta(x,y)-[\pi,\beta](x,y)+[\Phi_\Gamma,\beta](x,y).
\eqno(11.5)
$$ 
Therefore, imposing that $\beta(x,y)$ satisfies (11.4), we obtain the further 
conditions
$$
[f,\beta(x,y)]=0,\qquad q_\pi\beta(x,y)=0,
\eqno(11.6)
$$
for all $f$ in $\varsigma(\goth h)$, where $q_\pi$ is defined by (2.9) with 
$\varpi$ substituted by $\pi$. Note that the first condition (11.6) coincides 
with the first condition (6.12). When $\pi^{ij}=\varpi^{ij}$, (11.6)
reduces to (6.12) and no further restriction is implied by (11.4).
In general, imposing (6.12) and (11.6) simultaneously is rather restrictive 
and only trivial solutions of these conditions are available.

Suppose ${\cal O}=\sigma(x,\tilde X)$ is of the form (6.6). Then, 
$\sigma(x,\tilde X)$ obeys (6.16). Using (10.11), one computes
$$
w_\pi\sigma(x,\tilde X)=
d\big(\sigma(x,\tilde X)-k_\pi\sigma(x,\tilde X)\big)
+k_\pi d_M\sigma(x,\tilde X),
\eqno(11.7)
$$ 
where $k_\pi$ is the degree $0$ derivation defined by (6.8) with 
$\varpi$ substituted by $\pi$. 
Therefore, imposing that $\sigma(x,\tilde X)$ satisfies (11.4), we get 
the further condition
$$
k_\pi d_M\sigma(x,\tilde X)=0.
\eqno(11.8)
$$ 
When $\pi^{ij}=\varpi^{ij}$, (11.8) reduces to (6.16) and no further 
restriction is implied by (11.4). (6.16) and (11.8) are simultaneously
solved by all closed $p$--forms $\sigma_{i_1\cdots i_p}$ of $M$. However, 
non trivial observables are yielded only for $p=0,~1,~2$.

\vskip .4cm
{\bf 12. Discussion and Examples}
\vskip .4cm
\par

In this final section, we illustrate the formal analysis worked out above 
by providing a few examples of manifolds $M$ endowed with a pair of 
$2$--vectors $\varpi^{ij}$, $\pi^{ij}$ satisfying (2.1), (9.2), (9.3). 
For convenience, we write the $2$--vector $\pi^{ij}$ as 
$$
\pi^{ij}=\varpi^{ij}+\vartheta^{ij},
\eqno(12.1)
$$
where $\vartheta^{ij}$ is a $2$--vector satisfying (9.2), (9.3) with 
$\pi^{ij}$ replaced by $\vartheta^{ij}$. 
Hamilton actions $\varsigma$ of a finite dimensional Lie algebra 
$\goth h$ on $M$ satisfying (9.4) are most efficiently
constructed as follows. One chooses a finite set of linearly independent 
functions of $\Cas_\pi(M)$ and defines $\goth h$ to be the Lie algebra 
spanned by these functions under Poisson brackets, so that $\varsigma$
becomes simply the identity map. 
In what follows, we follow closely the methodology of ref. \ref{17}.

\vskip .3cm
{\it 2--Dimensional Poisson Spaces}.
\vskip .3cm

Let $M$ be a $2$--dimensional manifold. We equip $M$ with an auxiliary
metric $g_{ij}$. Any $2$--vector $\zeta^{ij}$ can be written as
$$
\zeta^{ij}=\epsilon^{ij}\alpha,
\eqno(12.2)
$$
for some function $\alpha$, where $\epsilon^{ij}$ is the Levi--Civita 
$2$--vector associated to $g_{ij}$. Let $\varpi^{ij}$, $\vartheta^{ij}$
be two $2$--vectors and let $\mu$, $\nu$ be the corresponding 
functions in the representation (12.2). Then $\varpi^{ij}$, $\vartheta^{ij}$ 
automatically are Schouten commuting Poisson $2$--vectors, irrespective of the 
specific form of $\mu$, $\nu$. A function $f$ on $M$ belongs to 
$\Cas_\pi(M)$, if and only if 
$$
(\mu+\nu)\partial_if=0.
\eqno(12.3)
$$
So, $f$ is constant in the open subsets of $M$ where the sum $\mu+\nu$ is
non vanishing and arbitrary in the open subsets of $M$ where $\mu+\nu$
vanishes. Since $\{f,g\}=\mu \epsilon^{ij}\partial_if\partial_jg$, 
$\Cas_\pi(M)$ is a generally non Abelian Poisson subalgebra of 
$\Fun(M)$.

\vskip .3cm
{\it 3--Dimensional Poisson Spaces}.
\vskip .3cm

Let $M$ be a $3$--dimensional manifold. We equip $M$ with an auxiliary
metric $g_{ij}$. Any $2$--vector $\zeta^{ij}$ can be written as
$$
\zeta^{ij}=\epsilon^{ijk}\alpha_k,
\eqno(12.4)
$$
for some $1$--form $\alpha_i$, where $\epsilon^{ijk}$ is the Levi--Civita 
$3$--vector associated to $g_{ij}$. Let $\varpi^{ij}$, $\vartheta^{ij}$
be two $2$--vectors and let $\mu_i$, $\nu_i$ be the corresponding 
$1$--forms in the representation (12.4). Then, $\varpi^{ij}$, $\vartheta^{ij}$
are Schouten commuting Poisson $2$--vectors, if and only if
$$
\epsilon^{ijk}\mu_i\nabla_j\mu_k=0, 
\quad\epsilon^{ijk}(\mu_i\nabla_j\nu_k+\nu_i\nabla_j\mu_k)=0,
\quad  \epsilon^{ijk}\nu_i\nabla_j\nu_k=0,
\eqno(12.5)
$$
where $\nabla_i$ is the Riemannian connection of $g_{ij}$.
It is known that the first and third condition have the local solution
$$
\mu_i=u\partial_ip, \qquad \nu_i=v\partial_iq,
\eqno(12.6)
$$
where $u$, $v$, $p$, $q$ are certain local functions \ref{17}. The remaining 
condition can then be cast as
$$
uv\epsilon^{ijk}\partial_i\ln\big(u/v\big)\partial_jp\partial_kq=0.
\eqno(12.7)
$$
A function $f$ on $M$ belongs to $\Cas_\pi(M)$, if and only if 
$$
\epsilon^{ijk}(\mu+\nu)_j\partial_kf=0,
\eqno(12.8)
$$
or, on account of (12.6),
$$
\epsilon^{ijk}(u\partial_jp+v\partial_jq)\partial_kf=0.
\eqno(12.9)
$$
By (12.8), if the $1$--form $\mu+\nu$ vanishes at most in the complement of 
an open dense set, then, at least locally, $\partial_if=k_f(\mu+\nu)_i$ 
for some function $k_f$. In that case, as it easy to see from the relation
$\{f,g\}=\epsilon^{ijk}\mu_i\partial_jf\partial_kg$, $\Cas_\pi(M)$ is 
an Abelian Poisson subalgebra of $\Fun(M)$. 
For instance, one may consider $M=\Bbb R^3$ equipped with
the Schouten commuting Poisson $2$--vectors $\varpi^{ij}$, $\vartheta^{ij}$
corresponding to the compatible Poisson structures
$$
\eqalignno{&\hbox to 4.5cm{$\{x_1,x_2\}\!\!\hphantom{{}_\varpi}=x_3$,\hfill}
\hbox to 3.0cm{$\{x_2,x_3\}\!\!\hphantom{{}_\varpi}=x_1$,\hfill}
\hbox to 3.0cm{$\{x_3,x_1\}\!\!\hphantom{{}_\varpi}=x_2$,\hfill}
\vphantom{\Big[}&(12.10)\cr
&\hbox to 4.5cm{$\{x_1,x_2\}_\vartheta
={1\over 2}-\big(x_3+{1\over 2})^2$,\hfill}
\hbox to 3.0cm{$\{x_2,x_3\}_\vartheta=0$,\hfill}
\hbox to 3.0cm{$\{x_3,x_1\}_\vartheta=0$.\hfill}
\vphantom{\Big[}&(12.11)\cr}
$$
The resulting Poisson $2$--vector $\pi^{ij}$ appears in the Poisson sigma 
model describing $2$--dimensional Euclidean $R^2$ gravity with 
cosmological constant \ref{4}.
A solution of eq. (12.8) is 
$$
f(x_1,x_2,x_3)=\hbox{$1\over 2$}\big(x_1{}^2+x_2{}^2\big)
-\hbox{$1\over 3$}x_3\big(x_3{}^2-\hbox{$3\over 4$}\big).
\eqno(12.12)
$$
As another example, one may consider $M=\Bbb R^2\times\Bbb S^1$ with
the Schouten commuting Poisson $2$--vectors $\varpi^{ij}$, $\vartheta^{ij}$
defined by the compatible Poisson structures
$$
\eqalignno{&\hbox to 3.0cm{$\{x_1,x_2\}\!\!\hphantom{{}_\varpi}=0$,\hfill}
\hbox to 4.0cm{$\{x_1,\varphi\}\!\!\hphantom{{}_\varpi}=0$,\hfill}
\hbox to 4.0cm{$\{x_2,\varphi\}\!\!\hphantom{{}_\varpi}=P(x_1,x_2)$,\hfill}
\vphantom{\Big[}&(12.13)\cr
&\hbox to 3.0cm{$\{x_1,x_2\}_\vartheta=0$,\hfill}
\hbox to 4.0cm{$\{x_1,\varphi\}_\vartheta=-Q(x_1,x_2)$,\hfill}
\hbox to 4.0cm{$\{x_2,\varphi\}_\vartheta=0$,\hfill}
\vphantom{\Big[}&(12.14)\cr}
$$
where $P(x_1,x_2)$, $Q(x_1,x_2)$ are certain functions
In this case, eq. (12.8) reduces to
$$
P(x_1,x_2)\partial_{x_2}f-Q(x_1,x_2)\partial_{x_1}f=0,
\quad P(x_1,x_2)\partial_\varphi f=Q(x_1,x_2)\partial_\varphi f=0.
\eqno(12.15)
$$
In the generic situation, $\partial_\varphi f=0$ and the first equation can be 
treated with standard analytical techniques.

\vskip .3cm
{\it 4--Dimensional Poisson Spaces}.
\vskip .3cm

Let $M$ be a $4$--dimensional manifold. We equip $M$ with an auxiliary
metric $g_{ij}$. Any $2$--vector $\zeta^{ij}$ can be written as
$$
\zeta^{ij}=\hbox{$1\over 2$}\epsilon^{ijkl}\alpha_{kl},\vphantom{\Big[}
\eqno(12.16)
$$
for some $2$--form $\alpha_{ij}$, where $\epsilon^{ijkl}$ is the Levi--Civita 
$4$--vector associated to $g_{ij}$. Let $\varpi^{ij}$, $\vartheta^{ij}$
be two $2$--vectors and let $\mu_{ij}$, $\nu_{ij}$ be the corresponding 
$2$--forms in the representation (12.16). Then, $\varpi^{ij}$, $\vartheta^{ij}$
are Schouten commuting Poisson $2$--vectors, if and only if
$$
\epsilon^{jklm}\mu_{jk}\nabla_l\mu_{mi}=0, 
\quad\epsilon^{jklm}(\mu_{jk}\nabla_l\nu_{mi}+\nu_{jk}\nabla_l\mu_{mi})=0,
\quad\epsilon^{jklm}\nu_{jk}\nabla_l\nu_{mi}=0,\vphantom{\Big[}
\eqno(12.17)
$$
where again $\nabla_i$ is the Riemannian connection of $g_{ij}$.
If one restricts oneself to degenerate Poisson $2$--vectors, i. e. with
everywhere vanishing determinant, it is known that the first and third 
condition have the local solution
$$
\mu_{ij}=u(\partial_ip\partial_jq-\partial_jp\partial_iq), 
\qquad \nu_{ij}=v(\partial_ir\partial_js-\partial_jr\partial_is),
\vphantom{\Big[}
\eqno(12.18)
$$
where $u$, $v$, $p$, $q$, $r$, $s$ are certain local functions \ref{17}. 
The remaining condition can then be cast as
$$
\eqalignno{
\epsilon^{jklm}\Big[&\,u\partial_jp\partial_kq\big(
\partial_mr\nabla_l(v\partial_is)
-\partial_ms\nabla_l(v\partial_ir)\big)\vphantom{\Big[}&(12.19)\cr
+&\,v\partial_jr\partial_ks\big(
\partial_mp\nabla_l(u\partial_iq)
-\partial_mq\nabla_l(u\partial_ip)\big)\Big]=0.\vphantom{\Big[}&\cr}
$$
A function $f$ on $M$ belongs to $\Cas_\pi(M)$, if and only if 
$$
\epsilon^{ijkl}(\mu+\nu)_{jk}\partial_lf=0,
\eqno(12.20)
$$
or, on account of (12.18),
$$
\epsilon^{ijkl}(u\partial_jp\partial_kq+v\partial_jr\partial_ks)\partial_lf=0.
\eqno(12.21)
$$
There is not much that can be said in general on the solution of this equation.
As an example, one can consider $M=\Bbb R^3\times \Bbb R$ equipped with
the Schouten commuting Poisson $2$--vectors $\varpi^{ij}$, $\vartheta^{ij}$
corresponding to the compatible Poisson structures
$$
\eqalignno{&\hbox to 4.5cm{$\{x_i,x_j\}\!\!\hphantom{{}_\varpi}=
\sum_{k=1}^3\varepsilon_{ijk}x_k y$,\hfill}
\hbox to 5.5cm{$\{x_i,y\}\!\!\hphantom{{}_\varpi}=0$,\hfill}
\vphantom{\Big[}&(12.22)\cr
&\hbox to 4.5cm{$\{x_i,x_j\}_\vartheta=0$,\hfill}
\hbox to 5.5cm{$\{x_i,y\}_\vartheta
=\sum_{j,k=1}^3\varepsilon_{ijk}(a_j-a_k)x_jx_k$,\hfill}
\vphantom{\Big[}&(12.23)\cr}
$$
where $\varepsilon_{ijk}$ is $3$--dimensional totally antisymmetric symbol
and the $a_i$ are real numbers.
The resulting Poisson $2$--vector $\pi^{ij}$ is that of the famous Sklyanin 
Poisson structure \ref{18}. Eq. (12.20) is solved by
$$
f_1(x_1,x_2,x_3,y)=\hbox{$1\over 2$}\sum_{i=1}^3a_ix_i{}^2
-\hbox{$1\over 4$}y^2,
\quad f_2(x_1,x_2,x_3,y)=\hbox{$1\over 2$}\sum_{i=1}^3x_i{}^2.
\eqno(12.24)
$$
$f_2$ is a common Casimir function of both $\varpi^{ij}$ and $\vartheta^{ij}$
and so is not of any use.

\vskip .3cm
{\it Affine Lie--Poisson spaces}.
\vskip .3cm

We consider $M=\Bbb R^n$ with the $2$--vectors
$$
\varpi^{ij}(x)=c^{ij}{}_kx^k, \quad \vartheta^{ij}(x)=a^{ij},
\eqno(12.25)
$$
where the constants $c^{ij}{}_k$, $a^{ij}$ satisfy
$$
\eqalignno{
&c^{ij}{}_mc^{mk}{}_l+c^{jk}{}_mc^{mi}{}_l+c^{ki}{}_mc^{mj}{}_l=0,
\vphantom{\Big[}&(12.26)\cr
&c^{ij}{}_ma^{mk}+c^{jk}{}_ma^{mi}+c^{ki}{}_ma^{mj}=0.
\vphantom{\Big[}&(12.27)\cr}
$$
As is well--known, (12.26), (12.27) state that $\Bbb R^{n\vee}$ is a Lie 
algebra with structure constants $c^{ij}{}_k$ and that $a^{ij}$ is a 
Chevalley--Eilenberg $2$--cocycle of $\Bbb R^{n\vee}$. $\varpi^{ij}$, 
$\vartheta^{ij}$ are Schouten commuting Poisson $2$--vectors. 
$\varpi^{ij}$ is usually called Kirillov--Kostant--Souriau Poisson structure
\ref{19-21}. A function $f$ on $M$ belongs to $\Cas_\pi(M)$, if and only if 
$$
(c^{ij}{}_kx^k+a^{ij})\partial_jf=0.
\eqno(12.28)
$$
An example is provided by $M=\Bbb R^4$ with the Poisson 
structures defined by 
$$
\eqalignno{&\hbox to 5.0cm{$\{x_0,x_i\}\!\!\hphantom{{}_\varpi}
=x_{i+1},\quad 1\leq i\leq 3$,\hfill}
\hbox to 5.0cm{$\{x_i,x_j\}\!\!\hphantom{{}_\varpi}=0,
\quad 1\leq i,j\leq 3$,\hfill}
\vphantom{\Big[}&(12.29)\cr
&\hbox to 5.0cm{$\{x_0,x_i\}_\vartheta=a\delta_{i,1},\quad 1\leq i\leq 3$,
\hfill}
\hbox to 5.0cm{$\{x_i,x_j\}_\vartheta=0,\quad 1\leq i,j\leq 3$,\hfill}
\vphantom{\Big[}&(12.30)\cr}
$$
where $a$ is a real number and $x_4=0$ by convention. A solution of (12.28)
is given by
$$
f(x_0,x_1,x_2,x_3)=\big(x_1x_3-\hbox{$1\over 2$}(x_2+a)^2\big)g(x_3),
\eqno(12.31)
$$
where $g$ is an arbitrary function.

\vskip .3cm
{\it Compact Poisson Riemannian Symmetric Spaces}.
\vskip .3cm

Let $M$ is a compact Riemannian symmetric space with metric $g_{ij}$. 
Then, if $\sigma_{ij}$, $\tau_{ij}$ are two harmonic $2$--forms, 
$$
\varpi^{ij}=g^{ik}g^{jl}\sigma_{kl}, \quad \vartheta^{ij}=g^{ik}g^{jl}\tau_{kl}
\eqno(12.32)
$$
are Schouten commuting Poisson $2$--vectors \ref{16}. Of course, this example 
is to be considered trivial unless the Betti number $b_2(M)\geq 2$,

\vskip.6cm
\par\noindent
{\bf Acknowledgments.} We are greatly indebted to R. Stora for providing 
his invaluable experience and relevant literature. We also thank the
referee of the paper for useful suggestions and improvements.
\vskip.6cm
\centerline{\bf REFERENCES}
\vskip.3cm

\item{[1]}
N.~Ikeda,
``Two-dimensional gravity and nonlinear gauge theory'',
Annals Phys.\  {\bf 235} (1994) 435
\item{} arXiv:hep-th/9312059.

\item{[2]}
P.~Schaller and T.~Strobl,
``Poisson structure induced (topological) field theories'',
Mod.\ Phys.\ Lett.\ A {\bf 9} (1994) 3129
\item{} arXiv:hep-th/9405110.

\item{[3]}
P.~Schaller and T.~Strobl,
``Poisson sigma models: A generalization of 2-d gravity Yang-Mills  systems'',
\item{} arXiv:hep-th/9411163.

\item{[4]}
P.~Schaller and T.~Strobl,
``A Brief Introduction to Poisson sigma-models'',
\item{} arXiv:hep-th/9507020.

\item{[5]}
M.~Kontsevich,
``Deformation quantization of Poisson manifolds'',
\item{} arXiv:q-alg/9709040.

\item{[6]}
A.~S.~Cattaneo and G.~Felder,
``A path integral approach to the Kontsevich quantization formula'',
Commun.\ Math.\ Phys.\  {\bf 212} (2000) 591
\item{} arXiv:math.qa/9902090.

\item{[7]}
A.~S.~Cattaneo and G.~Felder,
``Poisson sigma models and deformation quantization'',
Mod.\ Phys.\ Lett.\ A {\bf 16} (2001) 179
\item{} arXiv:hep-th/0102208.

\item{[8]}
A.~S.~Cattaneo and G.~Felder,
``On the AKSZ formulation of the Poisson sigma model'',
Lett.\ Math.\ Phys.\  {\bf 56} (2001) 163
\item{} arXiv:math.qa/0102108.

\item{[9]}
L.~Baulieu, A.~S.~Losev and N.~A.~Nekrasov,
``Target space symmetries in topological theories. I'''
\item{} arXiv:hep-th/0106042.

\item{[10]}
J.~Gomis, J.~Paris and S.~Samuel,
``Antibracket, antifields and gauge theory quantization'',
Phys.\ Rept.\  {\bf 259} (1995) 1
\item{} arXiv:hep-th/9412228.

\item{[11]}
S.~Cordes, G.~W.~Moore and S.~Ramgoolam,
``Lectures on 2-d Yang-Mills theory, equivariant cohomology and 
topological field theories'',
Nucl.\ Phys.\ Proc.\ Suppl.\  {\bf 41} (1995) 184
\item{} arXiv:hep-th/9411210.

\item{[12]} 
S.~Ouvry, R.~Stora and P.~van Baal, 
``On the Algebraic Characterization of Witten Topological Yang--Mills Theory'',
Phys.\ Lett.\ B {\bf 220} (1989) 159.

\item{[13]} 
V.~Mathai and D.~Quillen, 
``Superconnections, Thom Classes and Equivariant Differential Forms'', 
Topology {\bf 25} (1986) 85.

\item{[14]} M.~F.~Atiyah and L.~Jeffrey, 
``Topological Lagrangians and Cohomology'', 
J.\ Geom.\ Phys.\ {\bf 7} (1990) 119.

\item{[15]} 
W. Grueb, S. Halperin and R. Vanstone,
``Connections, Curvature and Cohomology'', vols. II and III, 
Academic Press, New York 1973.

\item{[16]}
I.~Vaisman,
``Lectures on the Geometry of Poisson Manifolds'',
Progress in Mathematics, vol. 118, Birkh\"auser Verlag, Basel 1994.

\item{[17]}
J.~Grabowski, G.~Marmo and A.~Perelomov,
``Poisson Structures: toward a Classification'',
Mod.\ Phys.\ Lett.\ A {\bf 8} (1993) 1719

\item{[18]}
E.~K.~Sklyanin,
``Some Algebraic Structures Connected With The Yang-Baxter Equation'',
Funct.\ Anal.\ Appl.\  {\bf 16} (1982) 263
[Funkt.\ Anal.\ Pril.\  {\bf 16N4} (1982) 27].

\item{[19]}
A.~A.~Kirillov,
``Unitary Representations of Nilpotent Lie Groups'',
Russian Math.\ Surveys {\bf 17} (1962) 53.

\item{[20]}
B.~Kostant,
``Orbits, Symplectic Structures and Representation Theory'',
Proc.\ U.S.--Japan Seminar of Diff.\ Geom.\ in Kyoto, (1966) 77.

\item{[21]}
J.~--M.~Souriau,
``Structure des Syst\`emes Dynamiques'',
Dunod, Paris 1970.

\bye